\documentclass[12pt, draftcls, onecolumn]{IEEEtran}

\usepackage[T1]{fontenc}
\usepackage{amssymb}
\usepackage{amsmath}
\usepackage{setspace}
\usepackage{algorithm}
\usepackage{algorithmic}

\usepackage{graphicx}
\usepackage{epstopdf}
\DeclareGraphicsExtensions{.eps}

\usepackage{subfigure}
\usepackage{cite}

\begin{document}
%
\title{A Rate Splitting Strategy for Massive MIMO with Imperfect CSIT}
\author{Mingbo~Dai, Bruno~Clerckx, David~Gesbert, Giuseppe~Caire
\thanks{M. Dai and B. Clerckx are with the Department of Electrical and Electronic Engineering, Imperial College London, UK, SW7 2AZ UK (e-mail: \{m.dai13, b.clerckx\}@imperial.ac.uk). B. Clerckx is also with the School of Electrical Engineering, Korea University, Seoul, Korea.
D. Gesbert are with the Mobile Communications Dept., EURECOM, 06560 Sophia Antipolis, France (e-mail: david.gesbert@eurecom.fr).
G. Caire is with the Department of Telecommunication Systems, Technical University of Berlin, Berlin, Germany (email: caire@tu-berlin.de).}
\thanks{This work was partially supported by the Seventh Framework Programme for Research of the European Commission under grant number HARP-318489. A shorter version of this work has been accepted by IEEE CAMAD 2015.}}

\maketitle

\begin{abstract}
In a multiuser MIMO broadcast channel, the rate performance is affected by the multiuser interference when the Channel State Information at the Transmitter (CSIT) is imperfect. To tackle the detrimental effect of the multiuser interference, a Rate-Splitting (RS) approach has been proposed recently, which splits one selected user's message into a common and a private part, and superimposes the common message on top of the private messages. The common message is drawn from a public codebook and should be decoded by all users. In this paper, we generalize the idea of RS into the large-scale array regime with imperfect CSIT. By further exploiting the channel second-order statistics, we propose a novel and general framework Hierarchical-Rate-Splitting (HRS) that is particularly suited to massive MIMO systems. HRS simultaneously transmits private messages intended to each user and two kinds of common messages that can be decoded by all users and by a subset of users, respectively. We analyse the asymptotic sum rate of RS and HRS and optimize the precoders of the common messages. A closed-form power allocation is derived which provides insights into the effects of system parameters. Finally, simulation results validate the significant sum rate gain of RS and HRS over various baselines. 

%
%
\end{abstract}


\IEEEpeerreviewmaketitle

\section{INTRODUCTION}
In MIMO wireless networks, the rate performance is affected by the multiuser interference when the Channel State Information at the Transmitter (CSIT) is imperfect \cite{bruno2013}. When the CSIT error variance $\tau^2$ decays with signal-to-noise ratio ($P$) as $O(P^{-\delta})$ for some constant $0 \le \delta < 1$, conventional multiuser broadcasting strategies (e.g., Zero-Forcing (ZF) beamforming) achieve the sum Degree-of-Freedom (DoF) $2\delta$ in a two-user MISO broadcast channel. Such a sum DoF performance reveals the bottleneck of a family of linear precoding schemes relying on imperfect CSIT as $\delta \rightarrow 0$. For example, the sum DoF of ZF is worse than single-user transmission strategy (e.g., TDMA) for $\delta < 0.5$ and becomes interference-limited for $\delta = 0$.

To address this issue, a Rate-Splitting (RS) approach was recently proposed\footnote{In \cite{yang2013}, the authors characterized the optimal DoF region of a two-user MISO broadcast channel with a mixture of imperfect current CSIT and perfect delayed CSIT. However, the corner points $(1, \delta)$ and $(\delta, 1)$ of the DoF region can be achieved with the Rate-Splitting approach, which does not exploit delayed CSIT and is applicable to the scenarios with only imperfect current CSIT.} \cite[Lemma 2]{yang2013}. Specifically, we can split one selected user's message (e.g., user $1$) into a common part $(s_c)$ and a private part $(s_1)$, where the common message is drawn from a public codebook and should be decoded by all users with zero error probability. The private message $(s_1)$ and the private messages $(s_{k|k \ne 1})$ intended to other users are transmitted via ZF beamforming using a fraction of the total power while the common message $(s_c)$ is superimposed on top of the ZF-precoded private messages using the residual power. At the receiver side, the common message is decoded by treating all the private messages as noise. After removing the decoded common message from the received signal by Successive Interference Cancellation (SIC), each user decodes their own private messages. With a proper power allocation, the sum DoF of RS is $1 + \delta$, which is strictly larger than the $2\delta$ that is achieved with ZF. It is worth mentioning that which user is selected to receive a common message in addition to a private message is irrelevant from a DoF perspective. Akin to rate-splitting strategies in the context of the Interference Channel (such as Han-Kobayashi scheme \cite{gam2012}), the common message is to be decoded by all users but is intended for a single user, which is highly different from the common message of the conventional multicasting techniques, where all users need the information contained in the common message \cite{luo2006}.

When the CSIT error variance $\tau^2$ is fixed and independent of $P$, linear precoding techniques with uniform power allocation lead to multiuser interference, which ultimately create a rate ceiling at high Signal-to-Noise-Ratio (SNR) \cite{jindal2006}. To circumvent this problem, one can adaptively tune the power allocation parameters among users as a function of SNR hence obtaining a single user transmission at extremely high SNR. Such an adaptive per-user power allocation bridges in a continuous manner the single-user mode and the multiuser mode. By contrast, RS can provide a rate performance beyond just operating the adaptive per-user power allocation. By optimizing the transmit beamforming and power allocation parameters for both RS and conventional multiuser broadcasting (BC) scheme, the RS approach shows significant sum rate gain over the conventional BC at finite SNR \cite{hamdi2015fir}. In the context of a two-user MISO broadcast channel with quantized CSIT, \cite{hao2015} has also confirmed the rate benefits provided by the transmission of a common message in a RS strategy over conventional multiuser BC schemes. Both the optimization method proposed in \cite{hamdi2015fir} and the analysis in \cite{hao2015} are hard to extend to multiuser massive MIMO systems.

Attaining accurate CSIT gets particularly challenging as the number of transmit antennas increases. Imperfect CSIT is therefore a major bottleneck for massive MIMO to realize its substantial spectral and energy efficiency benefits \cite{andrews2014, lulu2013}. In time division duplexing (TDD) systems, the downlink channels can be obtained by uplink training and channel reciprocity. CSIT might still be imperfect due to antenna miscalibration, imperfect channel estimation during the training phase and pilot contamination. In frequency division duplexing (FDD) systems, the downlink channels are estimated by the users from training phase and then fed back to the base station (BS) via uplink signaling. For a large number of transmit antennas, the downlink training represents a significant bottleneck and the corresponding feedback overhead is typically too large to afford. With limited feedback, it is practically infeasible to obtain good CSIT quality. Since massive MIMO is highly sensitive to CSIT quality, imperfect CSIT degrades the benefits of massive MIMO. In this paper, we will investigate the potential benefits of RS in the context of massive MIMO with imperfect CSIT.

When it comes to designing precoders on the basis of reduced CSI feedback, a two-tier precoder relying on both short- and long-term CSIT has been proposed by several authors \cite{bruno2008-1, bruno2008-2, li2010, lim2013, adhikary2013, Jaehyun2014, Kim2014, chen2014}. The dimensionality reduction offered by the two-tier precoder was shown to be very beneficial to multiuser MIMO in deployments with spatially correlated fading \cite{bruno2008-1, bruno2008-2}. This precoder structure also made its way to realistic systems as IEEE 802.16m and LTE-A \cite{li2010, lim2013}. When users are clustered into groups according to the similarity of their channel covariance matrices, \cite{adhikary2013} proposed a two-tier precoding approach to achieve massive-MIMO-like gains with highly reduced-dimensional CSIT. More precisely, the outer precoder controls inter-group interference based on long-term CSIT (the channel covariance matrices) while the inner precoder controls intra-group interference based on short-term effective channel (the channel concatenated with the outer precoder) with a reduced-dimension. The finding of \cite{adhikary2013} has been generalized into multi-polarized system \cite{Jaehyun2014}, where antenna polarization can be regarded as long-term CSI and used to further reduce the signaling overhead for CSIT acquisition. The work \cite{Kim2014} proposed a signal-to-leakage-plus-noise ratio (SLNR)-based outer precoder design and \cite{chen2014} developed a low complexity iterative algorithm to compute the outer precoder.

However, the system performance of the aforementioned two-tier precoding schemes are highly degraded by two limiting factors: inter-group and intra-group interference. When the eigen-subspaces spanned by the dominant eigenvectors of groups' spatial correlation matrices severely overlap, the outer precoder design may leak power (inter-group interference) to unintended groups. A typical example of overlapping eigen-subspace is that of users in different groups sharing common scatterers. Besides, randomly located users are not naturally partitioned into groups with the same covariance matrix. When user grouping techniques (e.g., K-mean clustering) are applied, the inter-group interference cannot be completely eliminated by the outer precoder \cite{nam2014}. In addition, the reduced-dimensional effective CSIT might be imperfect due to limited feedback, which leads to intra-group interference. To address these issues in the context of reduced CSI feedback, we develop a novel rate-splitting scheme that notably outperforms the conventional broadcasting schemes with two-tier precoder. The main contributions are listed as follows:

\begin{itemize}
  \item We generalize the idea of RS to massive MIMO deployments with imperfect CSIT. In fact, RS behaves as multiuser broadcasting in the low SNR regime, i.e., only private messages are transmitted. At high SNR, by transmitting a common message, the asymptotic multiplexing gain of RS amounts to that of single-user transmission. Meanwhile, RS exploits the rate benefits of multiuser broadcasting schemes by transmitting the private messages with a fraction of the total power. With the derived power allocation parameters, simulation results reveal that RS yields significant sum rate gain over conventional multiuser broadcasting and single-user transmission.

  \item In the event of spatially correlated channels, we propose a novel and general framework, called Hierarchical-Rate-Splitting (HRS), that is particularly suited to massive MIMO systems. By clustering users into groups based on their channel second-order statistics, the proposed HRS scheme exploits the benefits of spatially correlated channels and the two-tier precoding structure. Specifically, HRS is partitioned into an inner RS and an outer RS. Let us imagine each group as a single user, an outer RS tackles the inter-group interference by packing part of one selected user's message into a common codeword that can be decoded by all users. In the presence of multiple users per group, an inner RS copes with the intra-group interference by packing part of one selected user's message into a common codeword that can be decoded by users in that group. Note that the achievable rate of the common message is the minimum rate among users that decode this common message. In contrast to RS\footnote{For clarity, we refer to `RS' as the conventional one with one-tier precoder and `HRS' as the novel one with two-tier precoder.} where the only one common message should be decoded by all users, HRS offers more benefits by transmitting multiple common messages while the achievable rate of each inner common message is the minimum rate among a smaller number of users.

  \item RS and the proposed HRS scheme are analysed in the large-scale array regime. The precoder of each common message is designed to maximize the minimum rate of that common message experienced by each user. We also compute a closed-form adaptive power allocation for each message. With the derived power allocation, RS and HRS exhibit robustness w.r.t. CSIT error and/or eigen-subspaces overlap. Moreover, we quantify the sum rate gain of RS (HRS) over conventional multiuser broadcasting scheme with one-tier precoder (two-tier precoder), which offers insights into the effect of system parameters, e.g., SNR, CSIT quality, spatial correlation, number of users, etc.
\end{itemize}

\emph{Organization:} Section \ref{systemmod} introduces the system model. Section \ref{RS} discusses the RS transmission scheme and elaborates the precoder design, asymptotic rate performance and power allocation. In Section \ref{HRS}, we develop and then analyse a novel and general framework HRS. Section \ref{numresults} presents the numerical results and Section \ref{conclusion} concludes the paper.

Bold lower case and upper case letters denote vectors and matrices, respectively. The notations {\small$[\mathbf{X}]_i,[\mathbf{X}]_{i,j}, \mathbf{X}^T, \mathbf{X}^H, \text{tr}(\mathbf{X}), E(\mathbf{X})$} denote the $i$-th column, the entry in the $i$-th row and $j$-th column, the transpose, conjugate transpose, trace and expectation of a matrix {\small$\mathbf{X}$}. $\|\mathbf{x}\|$ represents the 2-norm of a vector.

\section{SYSTEM MODEL} \label{systemmod}
Consider a single cell FDD downlink system where the BS equipped with $M$ antennas transmits messages to $K (\le M)$ single-antenna users over a spatially correlated Rayleigh-fading channel. Consider a geometrical one-ring scattering model \cite{bruno2013}, the correlation between the channel coefficients of antennas $1 \le i, j \le M$ is given by

\begin{equation} \label{eq:correlation}
[\mathbf{R}_k]_{i,j} =  \frac{1}{2 \Delta_k} \int^{\theta_k + \Delta_k}_{\theta_k - \Delta_k} e^{-j \frac{2 \pi}{\lambda} \Psi(\alpha)(\mathbf{r}_i - \mathbf{r}_j) } d \alpha,
\end{equation}
where $\theta_k$ is the azimuth angle of user $k$ with respect to the orientation perpendicular to the array axis. $\Delta_k$ indicates the angular spread of departure to user $k$. $\Psi(\alpha) = [\cos(\alpha), \sin(\alpha) ]$ is the wave vector for a planar wave impinging with the angle of $\alpha$, $\lambda$ is the wavelength and $\mathbf{r}_i = [x_i, y_i]^T$ is the position vector of the $i$-th antenna. With the Karhunen-Loeve model, the downlink channel of user $k$ $\mathbf{h}_k \in \mathbb{C}^{M}$ is expressed as
\begin{equation} \label{eq:channel1}
\mathbf{h}_k =  \mathbf{U}_k \mathbf{\Lambda}^{\frac{1}{2}}_k \mathbf{g}_k,
\end{equation}
where $\mathbf{\Lambda}_k \in \mathbb{C}^{r_k \times r_k}$ is a diagonal matrix containing the non-zero eigenvalues of the spatial correlation matrix $\mathbf{R}_k$, and $\mathbf{U}_k \in \mathbb{C}^{M \times r_k}$ consists of the associated eigenvectors. The slowly-varying channel statistics $\mathbf{R}_k$ can be accurately obtained via a rate-limited backhaul link or via uplink-downlink reciprocity and is assumed perfectly known to both BS and users. We also assume block fading channel where $\mathbf{g}_k \in \mathbb{C}^{r_k}$ is static within a certain time slot but changes independently across different time slots and $\mathbf{g}_k$ has independent and identical distributed (i.i.d.) $\mathcal{CN} (0,1)$ entries.
For each channel use, linear precoding is employed at the BS to support simultaneous downlink transmissions to $K$ users. The received signals can be expressed as
\begin{equation} \label{eq:rx_ch}
\mathbf{y} = \mathbf{H}^H \mathbf{x} + \mathbf{n},
\end{equation}
where $\mathbf{x} \in \mathbb{C}^{M}$ is the linearly precoded signal vector subject to the transmit power constraint $\mathbb{E}[||\mathbf{x}||^2] \le P$, $\mathbf{H} = [\mathbf{h}_1, \cdots, \mathbf{h}_K]$ is the downlink channel matrix, $\mathbf{n} \sim \mathcal{CN} (\mathbf{0},\mathbf{I}_{K})$ is the additive white Gaussian noise (AWGN) vector and $\mathbf{y} \in \mathbb{C}^{K}$ is the received signal vector at the $K$ users.

\section{RATE-SPLITTING} \label{RS}
In this section, we introduce the RS transmission strategy and elaborate the precoder design, asymptotic rate performance as well as power allocation. The sum rate gain of RS over conventional multiuser broadcasting scheme will be quantified.

\subsection{CSIT Model} \label{RSCSITmodel}
Due to limited feedback (e.g., quantized feedback with a fixed number of quantization bits), only an imperfect channel estimate $\hat{\mathbf{h}}_k$ is available at the BS which is modeled as \cite{Wang2006}:
\vspace{-2pt}
\begin{equation} \label{eq:channel2}
\hat{\mathbf{h}}_k  = \mathbf{U}_k \mathbf{\Lambda}^{\frac{1}{2}}_k \, \hat{\mathbf{g}}_k = \mathbf{U}_k \mathbf{\Lambda}^{\frac{1}{2}}_k \Big(\sqrt{1 - \tau^2_k} \, \mathbf{g}_k + \tau_k \mathbf{z}_k \Big),
\end{equation}
where $\mathbf{z}_k$ has i.i.d. $\mathcal{CN} (0,1)$ entries independent of $\mathbf{g}_k$. $\tau_k \in [0, 1]$ indicates the quality of instantaneous CSIT for user $k$, i.e., $\tau_k = 0$ implies perfect CSIT whereas for $\tau_k = 1$ the CSIT estimate is completely uncorrelated with the true channel.

\subsection{Transmission Scheme} \label{RSmodel}
Let us firstly consider conventional linearly precoded multiuser broadcasting (BC) with one-tier precoder. The transmitted signal and received signal of user $k$ can be written as
\begin{equation} \label{eq:tx_sigbc}
\begin{array}{lcl}
\mathbf{x} &=& \mathbf{W} \mathbf{s} \;=\; \sum_{k = 1}^{K} \sqrt{P_{k}}  \mathbf{w}_{k} \, s_{k}, \vspace{0pt}\\
y_k &=& \sqrt{P_{k}} \, \mathbf{h}^H_{k} \mathbf{w}_{k} \, s_{k} + \underbrace{\sum_{j \ne k}^{K} \, \sqrt{P_{j}} \, \mathbf{h}^H_{k} \mathbf{w}_{j} \, s_{j}}_{\text{multiuser interference}} + n_k,
\end{array}
\end{equation}
where $\mathbf{s} = [s_1, \cdots, s_K]^T \in \mathbb{C}^K$ is the data vector intended for the $K$ users. Based on the imperfect channel estimate {\small $\hat{\mathbf{H}} = [\hat{\mathbf{h}}_1, \cdots, \hat{\mathbf{h}}_K]$}, the $M \times K$ precoder {\small$\mathbf{W} = [\mathbf{w}_1, \cdots, \mathbf{w}_K]$} is designed as ZF or Regularized-ZF (RZF). In the presence of imperfect CSIT with fixed error variance, the sum rate of BC with uniform power allocation is multiuser interference-limited at high SNR. In order to tackle the interference, one can adaptively schedule a smaller number of users to transmit as the SNR increases, which boils down to TDMA at extremely high SNR. However, such an adaptive scheduling is computationally heavy for a large number of users.


In the RS scheme, the message intended to one selected user\footnote{As far as a fairness problem is concerned, RS can be applied to each user in a round robin manner.} (e.g., user $1$) is split into a common part $(s_c)$ and a private part $(s_1)$. The common message $(s_c)$ is drawn from a public codebook and can be decoded by all users with zero error probability. The private message $(s_1)$ and the private messages $(s_{k|k \ne 1})$ intended to other users are superimposed over the common message and then transmitted with linear precoding. Specifically, the transmitted signal of RS can be written as
\begin{equation} \label{eq:tx_sigjmb}
\begin{array}{lcl}
\mathbf{x} = \sqrt{P_{c}} \, \mathbf{w}_{c} \, s_{c} + \sum_{k = 1}^{K} \, \sqrt{P_{k}} \, \mathbf{w}_{k} \, s_{k},
\end{array}
\end{equation}
where $\mathbf{w}_{c}$ is the unit-norm precoding vector of the common message. We here perform a uniform\footnote{A per-user power allocation optimization can further enhance the rate performance of RS, as has been done in \cite{hamdi2015fir}. In this paper, we consider a uniform power allocation to the private messages for both RS and conventional BC.} power allocation for the private messages. The interest is on how to allocate power between the common and private messages. Hence, the powers allocated to each message are given by $P_c = P(1 - t)$ and $P_k = Pt/K$, where $t \in (0, \, 1]$ denotes the fraction of the total power that is allocated to the private messages. The decoding procedure is performed as follows. Each user decodes first the common message $s_c$ by treating all private messages as noise. After eliminating the decoded common message by SIC, each user decodes their own private messages. By plugging \eqref{eq:tx_sigjmb} into \eqref{eq:rx_ch}, the Signal-to-Interference-plus-Noise-Ratios (SINRs) of the common message and the private message of user $k$ are written as
\begin{eqnarray} \label{eq:rs_sinr1}
\gamma^{c}_{k} = \frac{P_{c} \, |\mathbf{h}^H_{k} \mathbf{w}_{c} |^2 }{\sum_{k=1}^K P_{k} \, |\mathbf{h}^H_{k} \mathbf{w}_{k} |^2 + 1},  \quad
\gamma^{p}_{k} = \frac{P_{k} \, |\mathbf{h}^H_{k} \mathbf{w}_{k} |^2 }{\sum_{j \ne k} P_{j} \, |\mathbf{h}^H_{k} \mathbf{w}_{j} |^2 + 1}.
\end{eqnarray}

Under Gaussian signaling, the achievable rate of the common message is given as {\small$R^{RS}_{c} =  \log_2 (1 + \gamma^{c})$}, where {\small$\gamma^{c} = \underset{k}{\min} \; \{\gamma^{c}_{k} \}$} ensuring that the common message can be successfully decoded by all users. The sum rate of the private messages is given as {\small$R^{RS}_{p} = \sum^{K}_{k=1} R^{RS}_{k} = \sum^{K}_{k=1} \log_2 (1 + \gamma^p_{k})$}. Then, the sum rate of RS is {\small$R^{RS}_{\scriptstyle{\text{sum}}} = R^{RS}_{c} + R^{RS}_{p}$}.

\subsection{Precoder Design} \label{RSprec}
Let us first review two basic results on large-dimensional random vectors \cite[Lemma 4]{hoydis2013} that will be useful afterwards. Let {\small$\mathbf{A} \in \mathbb{C}^{M \times M}$} be uniformly bounded spectral norm w.r.t. $M$ and {\small$\mathbf{x}, \mathbf{y} \sim \mathcal{CN}(\mathbf{0}, \mathbf{I}_M )$} be mutually independent of {\small$\mathbf{A}$}. Then, we have almost surely that
\begin{equation} \label{lemma1}
\frac{1}{M} \mathbf{x}^H \mathbf{A} \mathbf{y}  \overset{M \rightarrow \infty}{\longrightarrow} 0, \quad
\frac{1}{M} \mathbf{x}^H \mathbf{A} \mathbf{x} - \frac{1}{M} \text{tr} (\mathbf{A}) \overset{M \rightarrow \infty}{\longrightarrow} 0.
\end{equation}

The precoder ($\mathbf{w}_k$) of the private message for MISO BC has been investigated in \cite{Bjornson2014} assuming perfect CSIT. The structure of the optimal $\mathbf{w}_k$ is a generalization of RZF precoding. In the presence of imperfect CSIT, the optimal precoders of the private messages are still unknown in simple closed-form and have to be optimized numerically following e.g. \cite{hamdi2015fir}. The optimization is particularly complex in large scale systems. Nevertheless, building upon \cite{hao2015, adhikary2013, wagner2012}, RZF based on the channel estimates {\small$\hat{\mathbf{H}}$} would be a suitable strategy for the precoders of the private messages. Hence,
\begin{equation} \label{eq:rsprecoder}
\mathbf{W} = \xi \, \hat{\mathbf{M}} \hat{\mathbf{H}},
\end{equation}
where {\small$\hat{\mathbf{M}} = (\hat{\mathbf{H}} \hat{\mathbf{H}}^H + M \, \varepsilon \, \mathbf{I}_{M} )^{-1}$} and $\varepsilon = {K}/{(M P)}$. To satisfy the transmit power constraint, the normalization scalar is set as {\small$\xi^2 = {K}/{\text{tr}(\hat{\mathbf{H}}^H \hat{\mathbf{M}}^H \hat{\mathbf{M}} \hat{\mathbf{H}})}$}.

The precoder $\mathbf{w}_c$ is designed to maximize the achievable rate of the common message, i.e., {\small$\log_2 (1 + \gamma^{c})$} with {\small$\gamma^{c} = \underset{k}{\min} \; \{\gamma^{c}_{k} \}$}. From \eqref{lemma1}, different channel estimates become asymptotically orthogonal in the large-scale array regime \cite{xiang2014}. Thus, we are able to design the precoder $\mathbf{w}_{c}$ in the subspace {\small$\mathcal{S} = \text{Span}(\hat{\mathbf{H}})$}, i.e.,
\vspace{-1pt}
\begin{equation} \label{eq:precoderopt2}
\mathbf{w}_{c} = \sum_{k} a_{k} \hat{\mathbf{h}}_{k},
\end{equation}
which can be interpreted as a weighted matched beamforming (MBF). The corresponding optimization problem is formulated as
\begin{equation} \label{eq:precoderopt}
\begin{array}{lcl}
\mathcal{P}1: \; \underset{\mathbf{w}_{c} \in \mathcal{S}}{\text{max}} \;\; \underset{k}{\text{min}} \;\, \pi_{k} \cdot |\mathbf{h}^H_{k} \mathbf{w}_{c} |^2  \\
\, \text{s.t.} \quad \|\mathbf{w}_{c}\|^2 = 1,
\end{array}
\end{equation}
where $\pi_{k} = \frac{P_{c}}{\sum_{k=1}^K P_{k} \, |\mathbf{h}^H_{k} \mathbf{w}_{k} |^2 + 1 }$ and the optimal solution $\{a^*_{k}\}$ is shown in the following proposition.

$\textbf{Proposition 1:}$ As $M \rightarrow \infty$, the asymptotically optimal solution of problem $\mathcal{P}1$ is given by
\begin{equation} \label{eq:optval}
a^*_{k} = \frac{1}{\sqrt{M \cdot \sum^K_{j=1} \frac{\pi_{k} \, (1 - \tau^2_{k})}{\pi_{j} \, (1 - \tau^2_{j})}} }, \; \forall k .
\end{equation}
\begin{IEEEproof}
By dividing the objective function of \eqref{eq:precoderopt} by $M^2$ and plugging \eqref{eq:precoderopt2} into \eqref{eq:precoderopt}, the problem $\mathcal{P}1$ is equivalently transformed to $\mathcal{P}2$:
\begin{equation} \label{eq:precoderopt3}
\begin{array}{lcl}
\mathcal{P}2: \; \underset{a_{k}}{\text{max}} \;\; \underset{k}{\text{min}} \;\, \pi_{k} \, (1 - \tau^2_{k}) \cdot a^2_{k}  \\
\, \text{s.t.} \quad \sum_{k} a^2_{k} = \frac{1}{M}.
\end{array}
\end{equation}

From \cite[Lemma 2]{xiang2014}, the optimal solution of problem $\mathcal{P}2$ is obtained when all terms are equal, i.e., $\pi_{k} \, (1 - \tau^2_{k}) \cdot a^2_{k} = \pi_{j} \, (1 - \tau^2_{j}) \cdot a^2_{j}, \, \forall k \ne j$ and the optimal $\{a^*_{k}\}$ are given by \eqref{eq:optval}.
\end{IEEEproof}

\emph{Remark 1}: The optimal precoder $\mathbf{w}^*_{c}$ is given by \eqref{eq:precoderopt2} with $\{a^*_{k}\}$ in \eqref{eq:optval}, by which all the $K$ users experience the same SINR \eqref{eq:rs_sinr1} w.r.t. the common message. Specially, the equally weighted MBF with {\small$a^*_{k} = {1}/{\sqrt{M K}}$} is optimal when the condition $\pi_{k} \, (1 - \tau^2_{k}) = \pi_{j} \, (1 - \tau^2_{j}), \, \forall k \ne j$ is satisfied. Nevertheless, we employ {\small$a^*_{k} = {1}/{\sqrt{M K}}$} rather than \eqref{eq:optval} for arbitrary cases, in order to obtain a more insightful and tractable asymptotic sum rate expression in the sequel.

\subsection{Asymptotic Rate Analysis} \label{RSperf}
We shall omit the proof of the following asymptotic SINRs of RS, which are straightforwardly established based on the approach of \cite{wagner2012}.

$\textbf{Theorem 1:}$ As $M, K \rightarrow \infty$ with a fixed ratio $\eta = \frac{M}{K}$, the SINRs \eqref{eq:rs_sinr1} of RS asymptotically converge as
\begin{equation} \label{eq:rs_desnr1}
\gamma^{c}_{k}  - \gamma^{c,\circ}_{k} \overset{M \rightarrow \infty}{\longrightarrow} 0, \quad
\gamma^{p}_{k} - \gamma^{p,\circ}_{k} \overset{M \rightarrow \infty}{\longrightarrow} 0,
\end{equation}
almost surely, where
\begin{equation} \label{eq:rs_desnr2}
\gamma^{c,\circ}_{k} = \frac{P(1 - t) (1 - \tau_k^2) \eta} {\frac{Pt}{K} \left(\xi^{\circ}\right)^2 (\Upsilon^{\circ}_{k} \Omega_{k} + \Phi_k)+ 1},
\end{equation}
\begin{equation} \label{eq:rs_desnr3}
\gamma^{p,\circ}_{k} = \frac{\frac{Pt}{K} \left(\xi^{\circ}\right)^2 \Phi_k} {\frac{Pt}{K} \left(\xi^{\circ}\right)^2 \Upsilon^{\circ}_{k} \Omega_{k}+ 1},
\end{equation}
and
\begin{equation} \label{eq:rs_desub1}
\left(\xi^{\circ}\right)^2 = \frac{K}{\Psi^{\circ}}, \quad \Psi^{\circ} = \frac{1}{M} \sum^K_{j = 1} \frac{ m'_{j}} { (1 + m^{\circ}_{j})^2},
\end{equation}
\begin{equation} \label{eq:rs_desub2}
\Upsilon^{\circ}_k = \frac{1}{M} \sum_{j \ne k} \frac{ m'_{j,k}} { (1 + m^{\circ}_{j})^2}, \quad \Phi_k = \frac{(1 - \tau_k^2) (m^{\circ}_{k})^2 }{(1 + m^{\circ}_{k})^2},
\end{equation}
\begin{equation} \label{eq:rs_desub3}
\Omega_{k} = \frac{1 - \tau_k^2(1 - (1 + m^{\circ}_{k})^2)}{(1 + m^{\circ}_{k})^2},
\end{equation}
with $\mathbf{m}' = [m'_1, \cdots, m'_K]^T$ and $\mathbf{m}'_k = [m'_{1,k}, \cdots, m'_{K,k}]^T$ defined by
\begin{equation} \label{eq:rs_desub4}
\mathbf{m}' = (\mathbf{I}_K - \mathbf{J})^{-1} \mathbf{v}, \quad
\mathbf{m}'_k = (\mathbf{I}_K - \mathbf{J})^{-1} \mathbf{v}_k,
\end{equation}
where $\mathbf{J}, \mathbf{v}$ and $\mathbf{v}_k$ are given as
\begin{equation}
[\mathbf{J}]_{i,j} = \frac{\frac{1}{M} \text{Tr}(\mathbf{R}_i \mathbf{T} \mathbf{R}_j \mathbf{T} ) }{M(1 + m^{\circ}_{j})^2}, \hspace{140pt}
\end{equation} \vspace{-20pt}
\begin{equation} \label{eq:rs_desub5}
\begin{array}{lcl}
\mathbf{v}_k &=& \Big[\frac{1}{M} \text{Tr}(\mathbf{R}_1 \mathbf{T} \mathbf{R}_k \mathbf{T}), \cdots, \frac{1}{M} \text{Tr}(\mathbf{R}_K \mathbf{T} \mathbf{R}_k \mathbf{T})\Big]^T, \\
\mathbf{v} &=& \Big[\frac{1}{M} \text{Tr}(\mathbf{R}_1 \mathbf{T}^2), \cdots, \frac{1}{M} \text{Tr}(\mathbf{R}_K \mathbf{T}^2)\Big]^T,
\end{array}
\end{equation}
and with $m^{\circ}_{k}$ and $\mathbf{T}_g$ the unique solutions of
\begin{equation} \label{eq:rs_desub6}
\begin{array}{lcl}
m^{\circ}_{k} = \frac{1}{M} \text{Tr}\left(\mathbf{R}_{k} \mathbf{T} \right), \quad \mathbf{T} = \left( \frac{1}{M} \sum_{j = 1}^K \frac{\mathbf{R}_{j}}{1 + m^{\circ}_{j}} + \varepsilon \, \mathbf{I}_{M} \right)^{-1}.
\end{array}
\end{equation}

By applying the continuous mapping theorem \cite{bill2008}, it follows from \eqref{eq:rs_desnr1} that {\small$R^{RS}_k - R^{RS,\circ}_k \overset{M \rightarrow \infty}{\longrightarrow} 0$}, where {\small$R^{RS,\circ}_k = \log_2 (1 + \gamma^{p,\circ}_{k})$} and {\small$R^{RS}_c - R^{RS,\circ}_c \overset{M \rightarrow \infty}{\longrightarrow} 0$}, where {\small$R^{RS,\circ}_c = \log_2 (1 + \gamma^{c,\circ})$} with {\small$\gamma^{c,\circ} = \underset{k}{\text{min}} \; \{\gamma^{c,\circ}_k\}$}. The asymptotic sum rate of the private messages is {\small$R^{RS,\circ}_p = \sum^K_{k=1} \log_2 (1 + \gamma^{p,\circ}_{k})$} and it follows that {\small$\frac{1}{K} (R^{RS}_p - R^{RS,\circ}_p) \overset{M \rightarrow \infty}{\longrightarrow} 0$}. Then, an approximation {\small$R^{RS,\circ}_{\scriptstyle{\text{sum}}}$} of the sum rate of RS is obtained as:
\begin{equation} \label{eq:rs_sumratede}
R^{RS,\circ}_{\scriptstyle{\text{sum}}} = R^{RS,\circ}_c + R^{RS,\circ}_p.
\end{equation}

According to random matrix theory and following \cite{adhikary2013,Jaehyun2014,wagner2012}, the asymptotic SINR/rate approximations become more accurate for increasing number of transmit antennas. Meanwhile, the asymptotic approximations are feasible and tight for large but finite $M$, e.g., 64 antennas implemented in the typical prototype of massive MIMO \cite{Erik2014}. Moreover, simulation results suggest that the asymptotes remain effective even for small system dimension, e.g., $M = 16$ \cite{wagner2012}.

\subsection{Power Allocation} \label{RSpower}
The optimal power splitting ratio $t$ that maximizes \eqref{eq:rs_sumratede} can be obtained by the first-order (derivative) condition. However, the solution is rather complicated. In this paper, we compute a suboptimal but effective and insightful power allocation method by which RS considerably outperforms conventional multiuser broadcasting schemes. We denote the asymptotic sum rate of broadcasting scheme in \eqref{eq:tx_sigbc} with RZF and uniform power allocation by {\small$R^{RZF,\circ}_{\scriptstyle{\text{sum}}} = \sum^K_{k=1} \log_2 (1 + \gamma^{RZF,\circ}_{k})$} with {\small$\gamma^{RZF, \circ}_k = \gamma^{p, \circ}_k|_{t = 1}$}. We can write
\begin{equation} \label{eq:rs_pow1}
\gamma^{p,\circ}_k = \frac{\frac{Pt}{K} \left(\xi^{\circ}\right)^2 \Phi_k} {\frac{Pt}{K} \left(\xi^{\circ}\right)^2 \Upsilon^{\circ}_{k} \Omega_{k}+ 1} \le \gamma^{RZF, \circ}_k = \frac{\frac{P}{K} \left(\xi^{\circ}\right)^2 \Phi_k} {\frac{P}{K} \left(\xi^{\circ}\right)^2 \Upsilon^{\circ}_{k} \Omega_{k}+ 1},
\end{equation}
for $\forall t \in (0, \,1]$. The basic idea is to allocate a fraction $(t)$ of the total power to transmit the private messages of RS and achieve approximately the same sum rate as the conventional multiuser BC with full power. Then, using the remaining power to transmit the common message of RS enhances the sum rate. The sum rate gain of RS over BC with RZF is quantified by
\begin{equation} \label{eq:rsgain}
\begin{array}{lcl}
\Delta R^{RS,\circ} = R^{RS,\circ}_c + \sum^K_{k = 1} \Big(\log_2 \big(1 + \gamma^{p,\circ}_{k}\big) - \log_2 \big(1 + \gamma^{RZF,\circ}_{k} \big) \Big).
\end{array}
\end{equation}

$\textbf{Proposition 2:}$ The equality of \eqref{eq:rs_pow1} holds when the power splitting ratio $t$ is given by
\begin{equation} \label{eq:poweralcrs2}
t = \text{min} \; \Big\{ \frac{K}{P \Gamma}, \; 1 \Big\},
\end{equation}
where $\Gamma = \underset{k}{\text{min}} \, \Big\{ \Upsilon^{\circ}_{k}\, \tau_k^2 /\Psi^{\circ} \Big\}$ and the sum rate gain $\Delta R^{RS,\circ}$ at high SNR is lower bounded as
\begin{equation} \label{eq:rsgain2}
\Delta R^{RS, \circ}  \ge \log_2 (1 + \gamma^{c,\circ}) - \log_2 (e).
\end{equation}
\begin{IEEEproof}
See Appendix \ref{sec:prop2}.
\end{IEEEproof}

\emph{Remark 2}: We can get useful insights into the effects of system parameters on the power allocation and sum rate gain. For example, $t$ decreases as $\tau^2$ increases, i.e., the power allocated to the private messages is reduced as the channel quality becomes worse $(\tau^2 \rightarrow 1)$. Moreover, the power allocated to the private messages $Pt$ is fixed at high SNR, in order to place the sum rate of the private messages back to the non-interference-limited regime. By assigning the remaining power $P - Pt$ to a common message, the sum rate increases with the available transmit power. As $P$ increases, $t \rightarrow 0$ but $t \ne 0$, i.e., RS will never boil down to single-user transmission and always exploits the benefits of private messages at low to medium SNR. In addition, from \eqref{eq:rsgain2}, the rate loss {\small$\big(R^{RZF,\circ}_{\scriptstyle{\text{sum}}} - R^{RS,\circ}_p \big)$} based on power allocation \eqref{eq:poweralcrs2} is upper bounded by $\log_2 (e) \approx 1.44$ bps/Hz. Last but not least, as $K$ becomes larger, $t (\le 1)$ increases which deteriorates {\small$\gamma^{c,\circ}_{k} = \frac{M}{K} P(1 - t) c$} for some constant $c>0$ and further the sum rate gain \eqref{eq:rsgain2}\footnote{By imposing a common decodability to the common message, the achievable rate associated to it may be severely limited in the large number of users case. One fundamental and interesting question arises: what if we design the common message to be decoded by a subset of users? The achievable rate of the common message would become higher since it is the minimum among a smaller number of users. Meanwhile, the achievable rates of those users that do not decode the common message will be degraded because of the interference from the common message. The optimal subset design (tradeoff) is under exploration.}.

\section{HIERARCHICAL-RATE-SPLITTING} \label{HRS}
As discussed before, a large number of users $(K)$ degrades the rate benefits of RS. Moreover, the BS requires channel estimate $\hat{\mathbf{H}}$ with a large dimension to perform conventional multiuser broadcasting scheme (e.g., ZF/RZF) or RS. In addition, the RS approach can be applied to spatially uncorrelated or correlated channels. But in its current form, RS does not make use of the channel second-order statistics, if known to the transmitter. Motivated by these considerations, we propose a novel and general framework, denoted as Hierarchical-Rate-Splitting (HRS), that exploits the knowledge of spatial correlation matrices and two kinds of common messages to enhance the sum rate and alleviate the CSIT requirement as well as the effect of large $K$.

\subsection{Transmission Scheme} \label{HRSmodel}
Recently, multiuser broadcasting schemes with a two-tier precoder for FDD massive MIMO systems have been proposed to lessen CSIT requirement by exploiting the knowledge of spatial correlation matrix at the transmitter \cite{adhikary2013, Jaehyun2014, Kim2014, chen2014}. Since the human activity is usually confined in a small region, locations of users tend to be spatially clustered. We make the same assumption as \cite{adhikary2013} that $K$ users are partitioned into $G$ groups (e.g., via K-mean clustering) and that users in each group share the same spatial correlation matrix {\small$\mathbf{R}_g = \mathbf{U}_g \mathbf{\Lambda}_g \mathbf{U}^H_g$} with rank $r_g$. We let $K_g$ denote the number of users in group $g$ such that {\small$\sum^G_{g=1} K_g = K$}. The downlink channel of the $g$-th group is expressed as {\small$\mathbf{H}_g =[\mathbf{h}_{g1}, \cdots, \mathbf{h}_{gK_g}] =  \mathbf{U}_g \mathbf{\Lambda}^{{1}/{2}}_g \mathbf{G}_g$}, where the elements of {\small$\mathbf{G}_g$} are distributed with $\mathcal{CN} (0, 1)$. Then, the transmitted signal of conventional two-tier precoded (TTP) broadcasting scheme is expressed as
\begin{equation} \label{eq:tx_sigjsdm}
\mathbf{x} = \sum_{g = 1}^{G} \,  \mathbf{B}_{g}  \mathbf{W}_{g} \mathbf{P}_{g}\, \mathbf{s}_{g},
\end{equation}
where {\small$\mathbf{s}_{g} \in \mathbb{C}^{K_g}$} represents the data streams for the $g$-th group users. The outer precoder {\small$\mathbf{B}_g \in \mathbb{C}^{M \times b_g}$} is based on the long-term CSIT while the inner precoder {\small$\mathbf{W}_g \in \mathbb{C}^{b_g \times K_g}$} depends on the short-term effective channel {\small$\bar{\mathbf{H}}_g = \mathbf{B}^H_g \mathbf{H}_g$}. {\small$\mathbf{P}_g \in \mathbb{C}^{K_g \times K_g}$} is the diagonal power allocation matrix with {\small$\mathbf{P}_g = \sqrt{{P}/{K}} \cdot \mathbf{I}$}. Then, the received signal of the $k$-th user in g-th group is given by
\begin{equation} \label{eq:rx_sigjsdm}
y_{gk} = \sqrt{P_{gk}} \mathbf{h}^H_{gk} \mathbf{B}_{g}  \mathbf{w}_{gk} s_{gk} + \underbrace{\sum_{j \ne k}^{K_g}  \sqrt{P_{gj}} \mathbf{h}^H_{gk} \mathbf{B}_{g}  \mathbf{w}_{gj} s_{gj}}_{\text{intra-group interference}} + \underbrace{\sum_{l \ne g}^{G} \, \mathbf{h}^H_{gk} \mathbf{B}_{l}  \mathbf{W}_{l} \mathbf{P}_{l}\, \mathbf{s}_{l}}_{\text{inter-group interference}} + n_{gk},
\end{equation}
where {\small$\mathbf{w}_{gk} = [\mathbf{W}_g]_k$}. To eliminate the inter-group interference, the outer precoder is designed in the nullspace of the eigen-subspace spanned by the dominant eigenvectors of the other groups' spatial correlation matrices. However, the power attached to the
weak eigenmodes may leak out to other groups and create inter-group interference. Besides, the intra-group interference cannot be completely be removed due to imperfect CSIT (e.g., limited feedback). To eliminate the interference-limited behavior at high SNR, one can optimize the groups, the users in each group, etc, as a function of the total transmit power and CSIT quality. In general, such an optimization problem is quite complex.

By generalizing the philosophy of RS in Section \ref{RS}, we propose a HRS scheme that consists of an outer RS and an inner RS. By treating each group as a single user, an outer RS would tackle the inter-group interference by packing part of one user's message into a common codeword that can be decoded by all users. Likewise, an inner RS would cope with the intra-group interference by packing part of one user's message into a common codeword that can be decoded by multiple users in that group. The common messages are superimposed over the private messages and the transmitted signal of HRS can be written as
\begin{equation} \label{eq:tx_sig}
\mathbf{x} = \sqrt{P_{oc}} \mathbf{w}_{oc} \, s_{oc} + \sum_{g = 1}^{G} \, \mathbf{B}_{g} \left( \sqrt{P_{ic,g}} \mathbf{w}_{ic,g} \, s_{ic,g} + \sqrt{P_{gk}} \mathbf{W}_{g} \, \mathbf{s}_{g} \right),
\end{equation}
where $s_{ic,g}$ denotes the inner common message intended to $g$-th group while $s_{oc}$ denotes the outer common message intended to all users. $\mathbf{w}_{ic,g}$ and $\mathbf{w}_{oc}$ are the corresponding unit norm precoding vectors. Similarly to RS, a uniform power allocation is performed for the private messages and we mainly focus on how to allocate power between the common and private messages. Hence, let $\beta \in (0, 1]$ represent the fraction of the total power that is allocated to the group (inner common and private) messages. Within each group, $\alpha \in (0, 1]$ denotes the fraction of power given to the private messages. Then, the power allocated to each message is jointly determined by $\alpha$ and $\beta$, i.e., $P_{oc} = P (1 - \beta), P_{ic,g} = \frac{P \beta}{G} (1 - \alpha), P_{gk} = \frac{P \beta}{K}\alpha$. The decoding procedure is performed as follows. Each user sequentially decodes $s_{oc}$ and $s_{ic,g}$, then remove them from the received signal by SIC. The private message intended to each user can be independently decoded by treating all other private messages as noise. By plugging \eqref{eq:tx_sig} into \eqref{eq:rx_ch}, the SINRs of the common messages and the private message of user $k$ are written as
\begin{eqnarray} \label{eq:gc_sinr}
\gamma^{oc}_{gk} &=& \frac{P_{oc} \, |\mathbf{h}^H_{gk} \mathbf{w}_{oc} |^2 }{ IN_{gk} } \\ \gamma^{ic}_{gk} &=& \frac{P_{ic,g} \, |\mathbf{h}^H_{gk} \mathbf{B}_g  \mathbf{w}_{ic,g}|^2 }{IN_{gk} - P_{ic,g} \, |\mathbf{h}^H_{gk} \mathbf{B}_g  \mathbf{w}_{ic,g}|^2} \\ \label{eq:sinr_gk}
\gamma^p_{gk} &=& \frac{P_{gk} \, |\mathbf{h}^H_{gk} \mathbf{B}_g \mathbf{w}_{gk}|^2 }{ IN_{gk} - P_{ic,g} \, |\mathbf{h}^H_{gk} \mathbf{B}_g  \mathbf{w}_{ic,g}|^2 - P_{gk} \, |\mathbf{h}^H_{gk} \mathbf{B}_g \mathbf{w}_{gk}|^2 },
\end{eqnarray}
where
\begin{equation} \label{eq:sc_sinr2}
\begin{array}{lcl}
IN_{gk} = \sum^G_{l=1} P_{ic,l} \, |\mathbf{h}^H_{gk} \mathbf{B}_l \mathbf{w}_{ic,l} |^2 + \sum^G_{l=1} \sum^{K_g}_{j=1} P_{lj} \, |\mathbf{h}^H_{gk} \mathbf{B}_l \mathbf{w}_{lj}|^2 + 1.
\end{array}
\end{equation}

The achievable rate of the outer common message is given by {\small$R^{HRS}_{oc} = \log_2 (1 + \gamma^{oc})$} with {\small$\gamma^{oc} = \underset{g,k}{\min} \; \{\gamma^{oc}_{gk} \}$}. The sum rate of the inner common messages is given by {\small$R^{HRS}_{ic} = \sum^G_{g = 1} R^{HRS}_{ic,g} =  \sum^G_{g = 1} \log_2 (1 + \gamma^{ic}_g)$} with {\small$\gamma^{ic}_g = \underset{k}{\min} \; \{ \gamma^{ic}_{gk} \}$}. The sum rate of the private messages is given as {\small$R^{HRS}_{p} = \sum^{G}_{g=1} \sum^{K_g}_{k=1} R^{HRS}_{gk} = \sum^{G}_{g=1} \sum^{K_g}_{k=1} \log_2 (1 + \gamma^p_{gk})$}. Then, the sum rate of HRS is written as {\small$R^{HRS}_{\scriptstyle{\text{sum}}} = R^{HRS}_{oc} + R^{HRS}_{ic} + R^{HRS}_{p}$}.

\subsection{Precoder Design} \label{HRSprec}
In contrast to RS, HRS has only access to the channel covariance matrices and the effective channel estimates {\small$\hat{\bar{\mathbf{H}}}_g = \mathbf{B}^H_g \hat{\mathbf{H}}_g$} of dimension $b_g \times K_g$, where {\small$\hat{\mathbf{H}}_g = \mathbf{U}_g \mathbf{\Lambda}^{\frac{1}{2}}_g \, \hat{\mathbf{G}}_g = \mathbf{U}_g \mathbf{\Lambda}^{{1}/{2}}_g (\sqrt{1 - \tau^2_g} \, \mathbf{G}_g + \tau_g \mathbf{Z}_g )$} has dimension of $M \times K_g$. Based on long-term CSIT, the outer precoder {\small$\mathbf{B}_{g}$} is designed to eliminate the leakage to other groups. Denoting the number of dominant (most significant) eigenvalues of {\small$\mathbf{R}_g$} by $r^d_g$ and collecting the associated eigenvectors as {\small$\mathbf{U}^d_g \in \mathbb{C}^{M \times r^d_g}$}, we define {\small$\mathbf{U}_{-g} = [\mathbf{U}^d_1, \cdots, \mathbf{U}^d_{g-1}, \mathbf{U}^d_{g+1}, \cdots, \mathbf{U}^d_G ] \in \mathbb{C}^{M \times \sum_{l \ne g} r^d_l }$}. According to the singular value decomposition (SVD), we denote by {\small$\mathbf{E}^{(0)}_{-g}$} the left eigenvectors of {\small$\mathbf{U}_{-g}$} corresponding to the $(M - \sum_{l \ne g} r^d_l )$ vanishing singular values. To reduce the inter-group interference while enhancing the desired signal power, {\small$\mathbf{B}_g$} is designed by concatenating {\small$\mathbf{E}^{(0)}_{-g}$} with the dominant eigenmodes of the covariance matrix of the projected channel {\small$\tilde{\mathbf{H}}_g = (\mathbf{E}^{(0)}_{-g})^H \mathbf{H}_g$}. The covariance matrix is decomposed as {\small$\tilde{\mathbf{R}}_g = (\mathbf{E}^{(0)}_{-g})^H \mathbf{U}_{g} \mathbf{\Lambda}_{g} \mathbf{U}^H_{g} \mathbf{E}^{(0)}_{-g}= \mathbf{F}_{g} \tilde{\mathbf{\Lambda}}_{g} \mathbf{F}^H_{g}$}, where {\small$\mathbf{F}_{g}$} includes the eigenvectors of {\small$\tilde{\mathbf{R}}_g$}. Denote {\small$\mathbf{F}^{(1)}_{g}$} as the dominant $b_g$ eigenmodes and then {\small$\mathbf{B}_g$} is given by
\begin{equation} \label{eq:outerprecoder}
\mathbf{B}_g = \mathbf{E}^{(0)}_{-g} \mathbf{F}^{(1)}_{g}.
\end{equation}

The outer precoder $\mathbf{B}_g$ can be interpreted as being the $b_g$ dominant eigenmodes that are orthogonal to the subspace spanned by the dominant eigen-space of groups $l \ne g$. $b_g$ determines the dimension of the effective channel and should satisfy $K_g \le b_g \le M - \sum_{l \ne g} r^d_l$ and $b_g \le r^d_g$. $r^d_g (\le r_g)$ is a design parameter with a sum rank constraint $\sum^G_{g=1} r^d_g \le M$.

The inner precoder $\mathbf{W}_g$ can be designed as RZF, i.e.,
\begin{equation} \label{eq:innerprecoder}
\mathbf{W}_g = \xi_g \, \hat{\bar{\mathbf{M}}}_g \hat{\bar{\mathbf{H}}}_g,
\end{equation}
where {\small$\hat{\bar{\mathbf{M}}}_g = (\hat{\bar{\mathbf{H}}}_g \hat{\bar{\mathbf{H}}}^H_g + b_g \, \varepsilon \, \mathbf{I}_{b_g} )^{-1}$}. By following \cite{adhikary2013, Jaehyun2014, wagner2012}, the regularization parameter is set as $\varepsilon = {K}/{b P}$ which is equivalent to the MMSE linear filter. $b$ is give by $b = \sum^G_{g=1} b_g$. Then, the power normalization factor is {\small$\xi^2_g = {K_g}/{\text{tr}(\hat{\bar{\mathbf{H}}}^H_g \hat{\bar{\mathbf{M}}}^H_g \mathbf{B}^H_g \mathbf{B}_g \hat{\bar{\mathbf{M}}}_g \hat{\bar{\mathbf{H}}}_g)}$}.

The precoder $\mathbf{w}_{oc} \in \mathbb{C}^{M}$ aims to maximize the achievable rate of the outer common message {\small$\log_2 (1 + \gamma^{oc})$} based on the reduced-dimensional channel estimate {\small$\hat{\bar{\mathbf{H}}}_g \in \mathbb{C}^{b_g \times K_g}, \, \forall g$}. However, there exists a dimension mismatch between $\mathbf{w}_{oc}$ and {\small$\hat{\bar{\mathbf{H}}}_g$}. To address this problem, we first construct {\small$\check{\mathbf{H}}_g = \mathbf{B}_g \hat{\bar{\mathbf{H}}}_g \in \mathbb{C}^{M \times K_g}$} and {\small$\check{\mathbf{H}} = [\check{\mathbf{H}}_1, \cdots, \check{\mathbf{H}}_G] \in \mathbb{C}^{M \times K}$}. From \eqref{lemma1}, the columns of {\small$\check{\mathbf{H}}$} become orthogonal as $M \rightarrow \infty$ and we are able to design the precoder $\mathbf{w}_{oc}$ in the subspace {\small$\mathcal{S} = \text{Span}(\check{\mathbf{H}})$}. Following Proposition 1 and Remark 1 in RS, we design the precoder $\mathbf{w}_{oc}$ as an equally weighted MBF, i.e., $\mathbf{w}_{oc} = \xi_{oc} \sum_{k} \frac{\check{\mathbf{h}}_{k}}{\sqrt{M}}$, where $\check{\mathbf{h}}_{k} = [\check{\mathbf{H}}]_{k}$ and $\xi_{oc}$ normalizes $\mathbf{w}_{oc}$ to unit norm.

On the other hand, the optimization of the multiuser transmit precoding is generally a NP-hard problem. Thus, the optimal precoder of the inner common message $\mathbf{w}_{ic,g}$ that maximizes {\small$R^{HRS}_{ic}$} cannot be obtained efficiently. However, when the outer precoder fully eliminates the inter-group interference, $\mathbf{w}_{ic,g}$ can be equivalently designed to maximize {\small$R^{HRS}_{ic,g}$} within each group. Following Proposition 1 and Remark 1, we here design $\mathbf{w}_{ic,g}$ as an equally weighted MBF of the effective channel {\small$\hat{\bar{\mathbf{H}}}_g$}. Under further assumption that $K \rightarrow \infty$, we note that {\small$\hat{\bar{\mathbf{M}}}_g$} of the inner precoder {\small$\mathbf{W}_g( = \xi_g \, \hat{\bar{\mathbf{M}}}_g \hat{\bar{\mathbf{H}}}_g)$} can be approximated by an identity matrix. Hence, $\mathbf{w}_{ic,g}$ can be equivalently designed as an equally weighted MBF of {\small$\mathbf{W}_g$}, i.e.,$\mathbf{w}_{ic,g} = \zeta_{ic,g} \hat{\bar{\mathbf{q}}}_g$, where $\hat{\bar{\mathbf{q}}}_g =  \frac{1}{K_g}\sum^{K_g}_{k = 1} \mathbf{w}_{gk}$ and {\small$\zeta^2_{ic,g} = {1}/{(\hat{\bar{\mathbf{q}}}^H_g \mathbf{B}^H_g \mathbf{B}_g \hat{\bar{\mathbf{q}}}_g)}$}.

\subsection{Asymptotic Rate Analysis} \label{HRSperf}
We shall omit the proof of the asymptotic SINRs of HRS, which is directly established based on the approach of \cite{wagner2012}.

$\textbf{Theorem 2:}$ As $M, K, b \rightarrow \infty$ with fixed ratios $\frac{K}{M}$ and $\frac{b}{M}$, the SINRs of HRS in \eqref{eq:gc_sinr} and \eqref{eq:sinr_gk} asymptotically converge as
\begin{equation} \label{eq:desnr1}
\begin{array}{lcl}
\gamma^{oc}_{gk}  - \gamma^{oc,\circ}_{g} \overset{M \rightarrow \infty}{\longrightarrow} 0, \quad
\gamma^{ic}_{gk}  - \gamma^{ic,\circ}_{g} \overset{M \rightarrow \infty}{\longrightarrow} 0, \quad
\gamma^{p}_{gk} - \gamma^{p,\circ}_{g} \overset{M \rightarrow \infty}{\longrightarrow} 0,
\end{array}
\end{equation}
almost surely, where \vspace{-0pt}
\begin{equation} \label{eq:desnr2}
\gamma^{oc,\circ}_{g} = \frac{ \kappa_g P(1 - \beta) (1 - \tau_g^2)} {\beta \big(\sum_{l \neq g} (\xi^{\circ}_{l})^2 \Upsilon^{\circ}_{gl} + \big(\xi^{\circ}_{g} \big)^2 \Upsilon^{\circ}_{gg} \Omega_g + \frac{P}{K} \big(\xi^{\circ}_{g} \big)^2 \Phi_g \big) + 1},
\end{equation}
\begin{equation} \label{eq:desnr3}
\gamma^{ic,\circ}_{g} = \frac{\beta (1 - \alpha) \big(\xi^{\circ}_{g} \big)^2 \left(\Upsilon^{\circ}_{gg} \Omega_g + \frac{P}{K} \Phi_g \right) }{\beta \sum_{l \neq g} \left(\xi^{\circ}_{l}\right)^2 \Upsilon^{\circ}_{gl} + \beta \alpha \, \big(\xi^{\circ}_{g} \big)^2 (\Upsilon^{\circ}_{gg} \Omega_g + \frac{P}{K} \Phi_g ) + 1},
\end{equation}
\begin{equation} \label{eq:desnr4}
\gamma^{p,\circ}_{g} = \frac{\beta \alpha \frac{P}{K} \big(\xi^{\circ}_{g} \big)^2 \Phi_g}{\beta \sum_{l \neq g} \left(\xi^{\circ}_{l}\right)^2 \Upsilon^{\circ}_{gl} + \beta \alpha \big(\xi^{\circ}_{g} \big)^2 \Upsilon^{\circ}_{gg} \Omega_g + 1},
\end{equation}
with \vspace{-0pt}
\begin{equation} \label{eq:desub1}
\left(\xi^{\circ}_{g}\right)^2 = \frac{K_g}{\Psi^{\circ}_g}, \quad \Psi^{\circ}_g = \frac{K_g}{b_g} \frac{ m'_{g}} { (1 + m^{\circ}_{g})^2},
\end{equation}
\begin{equation} \label{eq:desub2}
\Phi_g = \frac{ (1 - \tau^2_g) (m^{\circ}_g)^2}{(1 + m^{\circ}_g)^2}, \quad \Upsilon^{\circ}_{gl} =\frac{P}{K} \frac{K_g}{b_g} \frac{m'_{gl}}{(1 + m^{\circ}_{l})^2},
\end{equation}
\begin{equation} \label{eq:desub32}
m'_{g} = \frac{\frac{1}{b_g} \text{Tr}(\bar{\mathbf{R}}_{gg} \mathbf{T}_g \mathbf{B}^H_g \mathbf{B}_g \mathbf{T}_g) }{1 - \frac{\frac{K_g}{b_g} \text{Tr}(\bar{\mathbf{R}}_{gg} \mathbf{T}_g \bar{\mathbf{R}}_{gg} \mathbf{T}_g) }{b_g \, (1 + m^{\circ}_{g})^2 } }, \quad
m'_{gl} = \frac{\frac{1}{b_g} \text{Tr}(\bar{\mathbf{R}}_{ll} \mathbf{T}_l \bar{\mathbf{R}}_{gl} \mathbf{T}_l) }{1 - \frac{\frac{K_g}{b_g} \text{Tr}(\bar{\mathbf{R}}_{ll} \mathbf{T}_l \bar{\mathbf{R}}_{ll} \mathbf{T}_l) }{b_g \, (1 + m^{\circ}_{l})^2 } },
\end{equation}
\begin{equation} \label{eq:desub4}
\Omega_g = \frac{K_g-1}{K_g}  \frac{(1 - \tau^2_g (1 - (1 + m^{\circ}_g)^2)) }{(1 + m^{\circ}_g)^2},
\end{equation}
\begin{equation} \label{eq:desub42}
\kappa_g = \frac{\text{Tr}(\bar{\mathbf{R}}_{gg})^2}{\sum^G_{l = 1} K_g \text{Tr}(\bar{\mathbf{R}}_{ll})}, \quad \bar{\mathbf{R}}_{gl} = \mathbf{B}^H_l \mathbf{R}_g \mathbf{B}_l, \; \forall g, l
\end{equation}
and $m^{\circ}_{g}$ and $\mathbf{T}_g$ the unique solutions of
\begin{equation} \label{eq:desub5}
m^{\circ}_{g} = \frac{1}{b_g} \text{Tr}\left(\bar{\mathbf{R}}_{gg} \mathbf{T}_g \right), \quad \mathbf{T}_g = \left( \frac{K_g}{b_g} \frac{\bar{\mathbf{R}}_{gg}}{1 + m^{\circ}_{g}} + \varepsilon \, \mathbf{I}_{b_g} \right)^{-1}.
\end{equation}

It follows from \eqref{eq:desnr1} that {\small$\frac{1}{K}(R^{HRS}_p - R^{HRS,\circ}_p) \overset{M \rightarrow \infty}{\longrightarrow} 0$} where {\small$R^{HRS,\circ}_p = \sum^G_{g=1} K_g \log_2 (1 + \gamma^{p, \circ}_{g})$, $\frac{1}{G}(R^{HRS}_{ic} - R^{HRS,\circ}_{ic}) \overset{M \rightarrow \infty}{\longrightarrow} 0$} where {\small$R^{HRS,\circ}_{ic} = \sum^G_{g=1} \log_2 (1 + \gamma^{ic, \circ}_{g})$}, and that {\small$R^{HRS}_{oc} - R^{HRS,\circ}_{oc} \overset{M \rightarrow \infty}{\longrightarrow} 0$} where {\small$R^{HRS,\circ}_{oc} = \log_2 (1 + \gamma^{oc, \circ})$} with {\small$\gamma^{oc,\circ} = \underset{g}{\text{min}} \; \{\gamma^{oc,\circ}_g\}$}. Then, an approximation {\small$R^{HRS,\circ}_{\scriptstyle{\text{sum}}}$} of the sum rate of HRS is obtained as:
\begin{equation} \label{eq:sumratede}
R^{HRS,\circ}_{\scriptstyle{\text{sum}}} = R^{HRS,\circ}_{oc} + R^{HRS,\circ}_{ic} + R^{HRS,\circ}_p.
\end{equation}

Likewise, the asymptotic sum rate of the conventional TTP in \eqref{eq:tx_sigjsdm} converges as {\small$(R^{TTP}_{\scriptstyle{\text{sum}}}- R^{TTP, \circ}_{\scriptstyle{\text{sum}}})/K \!\overset{M \rightarrow \infty}{\longrightarrow}\! 0$}, where {\small$R^{TTP, \circ}_{\scriptstyle{\text{sum}}} = \sum^G_{g=1} K_g \log_2 (1 + \gamma^{TTP, \circ}_{g})$} and
\begin{equation} \label{eq:snrjsdm}
\gamma^{TTP, \circ}_{g} = \frac{\frac{P}{K} \big(\xi^{\circ}_{g} \big)^2 \Phi_g}{\sum_{l \neq g} \left(\xi^{\circ}_{l}\right)^2 \Upsilon^{\circ}_{gl} + \big(\xi^{\circ}_{g} \big)^2 \Upsilon^{\circ}_{gg} \Omega_g + 1},
\end{equation}
and the first term in the denominator of \eqref{eq:snrjsdm} containing $\Upsilon^{\circ}_{gl} (\bar{\mathbf{R}}_{gl})$ denotes inter-group interference while the second term with $\Omega_g (\tau^2)$ refers to intra-group interference. The sum rate gain of HRS over conventional two-tier precoding BC is quantified by
\begin{equation} \label{eq:hrsgain}
\begin{array}{lcl}
\Delta R^{HRS,\circ} = R^{HRS,\circ}_{oc} + R^{HRS,\circ}_{ic} + \sum^G_{g = 1} K_g \Big(\log_2 \big(1 + \gamma^{p,\circ}_{g}\big) - \log_2 \big(1 + \gamma^{TTP,\circ}_{g} \big) \Big).
\end{array}
\end{equation}

\subsection{Power Allocation} \label{HRSpower}
Since $\alpha$ and $\beta$ are coupled in the SINR expressions \eqref{eq:desnr2} $\sim$ \eqref{eq:desnr4},  a closed-form and optimal solution that maximizes the sum rate of HRS {\small$R^{HRS,\circ}_{\scriptstyle{\text{sum}}}$} cannot be obtained in general. Following a similar philosophy as in \ref{RSpower}, we compute a closed-form suboptimal but effective power allocation method, by which the private messages of HRS are allocated a fraction of the total power and achieve nearly the same sum rate as the conventional broadcasting scheme with full power, i.e., {\small$R^{TTP,\circ}_{\scriptstyle{\text{sum}}} \approx R^{HRS,\circ}_p$}. Then, the remaining power is utilized to transmit the common messages and therefore enhance the sum rate. We can write
\begin{equation} \label{eq:poweralc1}
\gamma^{p,\circ}_g \le \gamma^{TTP, \circ}_g, \quad \forall g,
\end{equation}
for $\forall \alpha, \beta \in (0, 1]$. Consider two extreme cases: weak and strong inter-group interference. Based on \eqref{eq:snrjsdm}, the notation of `weak' implies that the inter-group interference is sufficiently small and therefore can be negligible, i.e., {\small$\Upsilon^{\circ}_{gl} \approx 0, \forall g \ne l$}. The sum rate {\small$R^{TTP,\circ}_{\scriptstyle{\text{sum}}}$} is limited by the intra-group interference due to imperfect CSIT. On the contrary, the notation of `strong' means that the inter-group interference dominates the rate performance, i.e., {\small$\sum_{l \neq g} \left(\xi^{\circ}_{l}\right)^2 \Upsilon^{\circ}_{gl} > \big(\xi^{\circ}_{g} \big)^2 \Upsilon^{\circ}_{gg}$}.

$\textbf{Proposition 3:}$ The equality of \eqref{eq:poweralc1} holds when the power splitting ratios $\alpha$, $\beta$ are given as
\begin{equation} \label{eq:poweralc2}
\beta = 1, \quad \alpha = \text{min} \; \Big\{ \frac{K_g}{P \cdot \Gamma_{IG}}, \; 1 \Big\}
\end{equation}
in the weak inter-group interference regime, and as
\begin{equation} \label{eq:poweralc22}
\beta = \text{min} \; \Big\{\frac{K}{P \cdot \Gamma_{OG} + K_g}, \; 1 \Big\}, \quad \alpha = 1
\end{equation}
in the strong inter-group interference regime, where
\begin{eqnarray} \label{eq:poweralc3}
\Gamma_{OG} &=& \underset{g}{\text{min}} \; \bigg\{ \sum_{l \ne g} \frac{K_g}{K} \frac{\text{tr}\big(\bar{\mathbf{R}}_{gl} \bar{\mathbf{R}}^{-1}_{ll}\big)}{\text{tr}\big(\bar{\mathbf{R}}^{-1}_{ll}\big)} \bigg\}, \\
\Gamma_{IG} &=& \underset{g}{\text{min}} \; \bigg\{ \frac{\tau_g^2}{K} \frac{b_g (K_g - 1)}{\text{tr}\big(\bar{\mathbf{R}}^{-1}_{gg}\big)} \bigg\}.
\end{eqnarray}
\begin{IEEEproof}
See Appendix \ref{sec:prop3}.
\end{IEEEproof}

When the inter-group interference is negligible, HRS becomes a set of parallel RS in each group, i.e., the outer common message is unnecessary. By contrast, when the inter-group interference is the dominant degrading factor, the inner common message transmission as well as the private messages transmission are inter-group interference limited. In this case, HRS boils down to RS (with reduced-dimensional CSIT). For the general inter-group interference case, finding simple closed-form $\alpha, \beta$ that guarantees a sum rate gain of HRS over two-tier precoding BC is challenging. Nevertheless, motivated by the design philosophy of power allocation in the extreme cases, we induce a threshold $\mu$ by which $\gamma^{p,\circ}_g = \mu \gamma^{TTP, \circ}_g$ (e.g., $\mu = 0.9$). By following the proof of Proposition 3, we compute the power allocation factors as follows: $\beta = \text{min} \; \Big\{\frac{K}{P \cdot (\Gamma_{OG} + \alpha \Gamma_{IG})}, \; 1 \Big\}, \; \alpha = \text{min} \; \Big\{ \frac{\mu (P \Gamma_{OG} + 1)}{P \cdot (\Gamma_{OG} + (1 - \mu) \Gamma_{IG}) + 1}, \; 1 \Big\}$. The threshold $\mu$ should be carefully designed for certain system setting.


From \eqref{eq:poweralc2} and \eqref{eq:poweralc22}, we have $\alpha = \beta = 1$ at low SNR and HRS becomes the conventional two-tier precoding BC, leading to {\small$\Delta R^{HRS,\circ} = 0$}. Namely, the effect of imperfect CSIT/overlapping eigen-subspaces on the sum rate of broadcasting private messages is negligible and thereby common message(s) is not needed. On the other hand, the rate performance of the conventional two-tier precoding BC schemes saturates at high SNR while HRS exploits a fraction of the total power ($\alpha < 1$ or $\beta < 1$) to transmit the common message(s) and enhance the sum rate.

$\textbf{Corollary 3.1:}$ With power allocation of Proposition 3, the sum rate gain $\Delta R^{HRS,\circ}$ at high SNR is lower bounded as:
\begin{equation} \label{eq:hrsgain1}
\Delta R^{HRS,\circ}  \ge \sum^G_{g=1} \big(\log_2 (1 + \gamma^{ic, \circ}_{g}) - \log_2 (e)\big),
\end{equation}
in the weak inter-group interference regime, and as
\begin{equation} \label{eq:hrsgain2}
\Delta R^{HRS,\circ}  \ge \log_2 (1 + \gamma^{oc, \circ}) - \log_2 (e),
\end{equation}
in the strong inter-group interference regime.
\begin{IEEEproof}
See Appendix \ref{sec:coro3}.
\end{IEEEproof}

\emph{Remark 3:} The following are some interpretations of Proposition 3 and Corollary 3.1.
\begin{itemize}
  \item \textbf{Power allocation to the private and common messages}: The intra-group power splitting ratio $(\alpha)$ decreases as $\tau^2$ increases. Namely, in order to alleviate intra-group interference, we should allocate less power to the private messages as the CSIT quality gets worse $(\tau^2 \rightarrow 1)$. Similarly, the inter-group power splitting ratio $(\beta)$ drops as the inter-group interference term $\Upsilon^{\circ}_{gl}, g \ne l$ becomes larger. From \eqref{eq:poweralc2} $\sim$ \eqref{eq:poweralc22}, the power distributed to the privates messages is an invariant of $P$ at high SNR:
      \begin{equation} \label{eq:remark11}
      \sum^G_{g=1}\sum^{K_g}_{k=1} P_{gk} = P \alpha \beta = \left\{
                    \begin{array}{lcl}
                        \frac{K}{\Gamma_{OG}} , \quad \text{if} \; \beta < 1 \\
                        \frac{K_g}{\Gamma_{IG}} , \quad \text{otherwise}
                    \end{array}
                    \right.
      \end{equation}
      which places the sum rate of private messages back into the non-interference-limited regime. Meanwhile, the power allocated to the common messages linearly increases with $P$ at high SNR.
  \item \textbf{Sum rate gain}: HRS exploits the extra power beyond saturation of conventional broadcasting schemes to transmit the common message(s), leading to a sum rate that increases with the available transmit power. In the weak inter-group interference regime, HRS becomes a set of parallel inner RS. Based on \eqref{eq:hrsgain1}, the sum rate gain {\small$\Delta R^{HRS,\circ}$} increases by $G$ bps/Hz for each 3 dB power increment at high SNR. By contrast, HRS boils down to RS in the strong inter-group interference regime and {\small$\Delta R^{HRS,\circ}$} increases by $1$ bps/Hz for each 3 dB power increment at high SNR.
\end{itemize}

\section{SIMULATION RESULTS} \label{numresults}
Numerical results are provided to validate the effectiveness of RS and HRS. Uniform circular array (UCA) with $M = 100$ isotropic antennas are equipped at the BS. Consider the transmit correlation model in \eqref{eq:correlation}, the antenna elements are equally spaced on a circle of radius $\lambda D$, for {\small$D = \frac{0.5}{\sqrt{(1-\cos(2 \pi/M))^2 + \sin(2 \pi/M)^2}}$}, leading to a minimum distance $\lambda/2$ between any two antennas.

\subsection{RS}

\begin{figure*}[t] \centering
\subfigure[]{ \label{fig_RS1} \includegraphics[width = 0.4\textwidth]{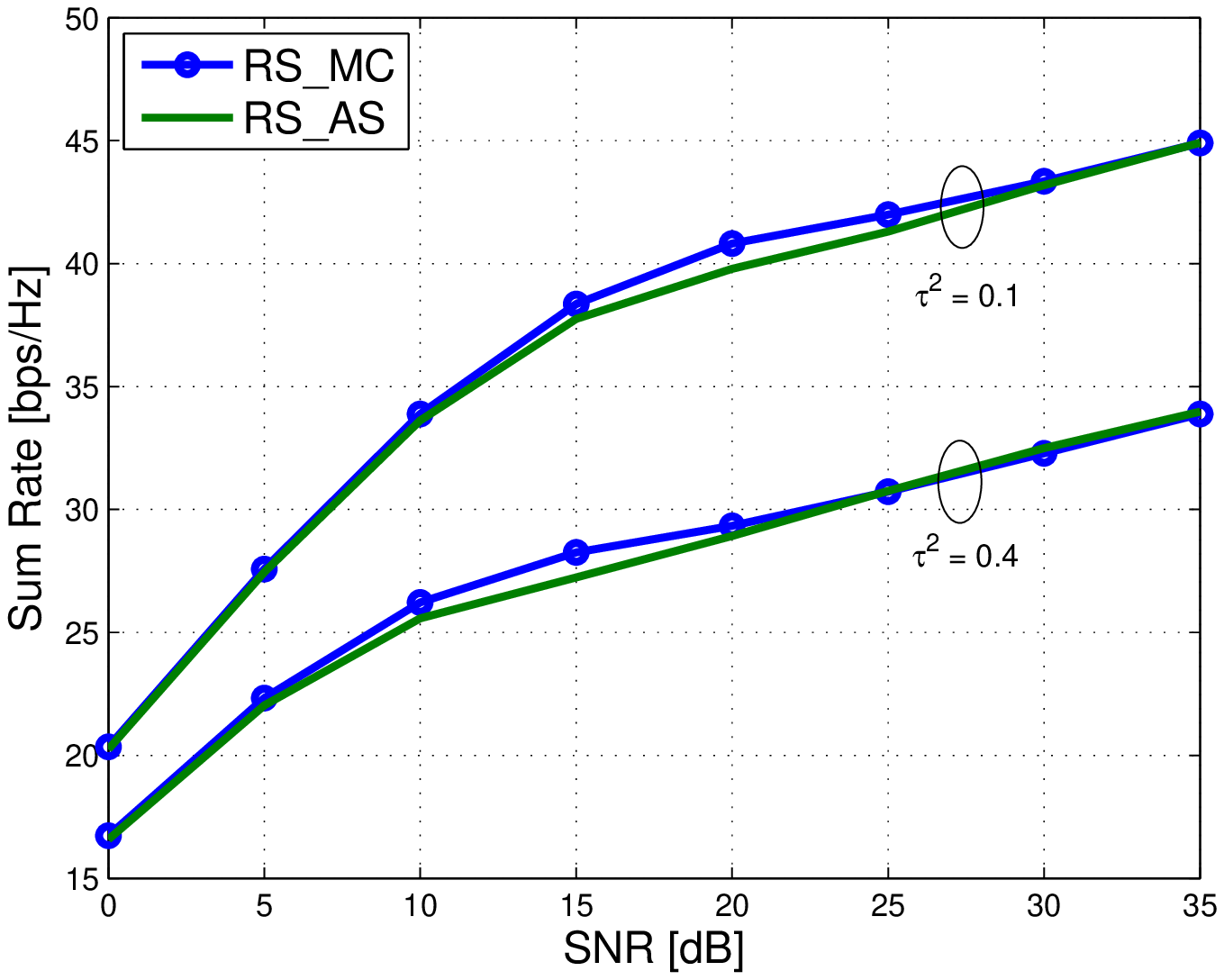}  } \hspace{20pt}
\subfigure[]{ \label{fig_RS2}  \includegraphics[width = 0.4\textwidth]{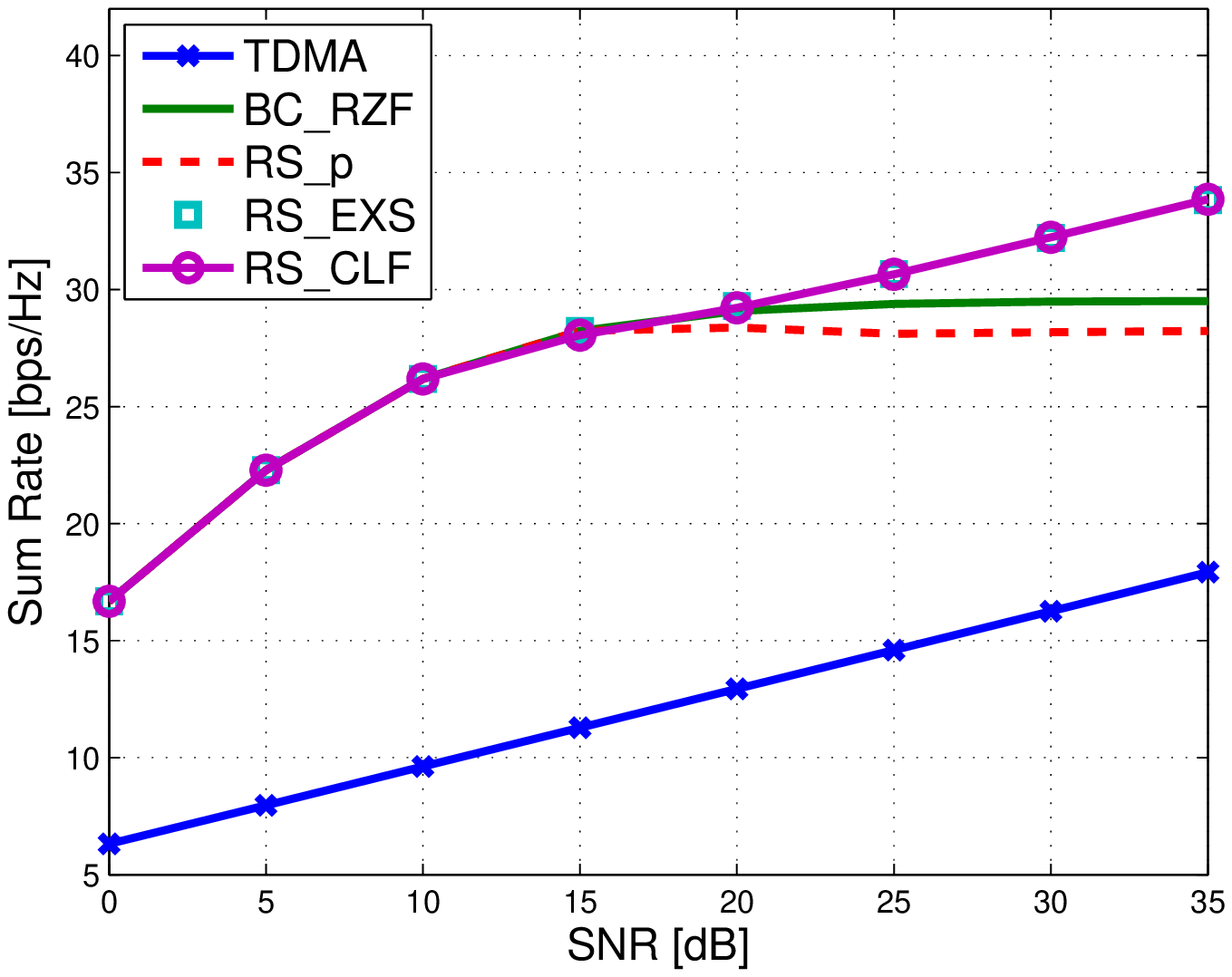} }
\caption{$K = 5$. (a) RS asymptote (AS) vs. RS Monte-Carlo (MC). (b) $\tau^2 = 0.4$, RS vs. BC with RZF.}
\end{figure*}

The users are assumed to be distributed uniformly at an azimuth angle $\theta_k = 2\pi k/K$ and angular spread $\Delta_k = \pi/6$. We compare RS with RZF-precoded multiuser broadcasting (BC\_RZF) \cite{wagner2012} and TDMA. Two types of RS are investigated: exhaustive search (RS\_EXS) and closed-form (RS\_CLF). Specifically, RS\_EXS performs a simulation-based exhaustive search with step 0.01 for the best power splitting ratio $t$. RS\_CLF allocates power by following Proposition 2.

In Fig. \ref{fig_RS1}, the asymptotic approximation {\small$R^{RS,\circ}_{\scriptstyle{\text{sum}}}$} in \eqref{eq:rs_sumratede} properly characterizes the sum rate of RS for various CSIT qualities and SNRs. From Fig.  \ref{fig_RS2}, we observe that RS\_CLF achieves almost the same sum rate as RS\_EXS. This verifies the effectiveness of the proposed power allocation strategy \eqref{eq:poweralcrs2}. Moreover, the multiplexing gain of RS is approaching 1 at high SNR. RS behaves as BC\_RZF in the low to medium SNR regime. At high SNR, the sum rate of RS linearly increases with the transmit power (dB) as TDMA. By contrast, BC\_RZF suffers from a rate ceiling due to imperfect CSIT. `RS\_p' denotes the sum rate of the private messages in RS transmission, which confirms the power allocation strategy in Section \ref{RSpower}. Namely, RS allocates a fraction of power to transmit the private messages so as to achieve approximately the same sum rate as BC\_RZF with full power. Then, using the remaining power to transmit the common message enhances the sum rate. In Fig. \ref{fig_RS2}, the rate loss {\small$R^{RZF}_{\scriptstyle{\text{sum}}} - R^{RS}_p$} is 1.3 bps/Hz at high SNR, which is lower than $\log_2(e)$ and agrees with Remark 2.

\begin{figure*}[t] \centering
\subfigure[]{ \label{fig_RS3} \includegraphics[width = 0.4\textwidth]{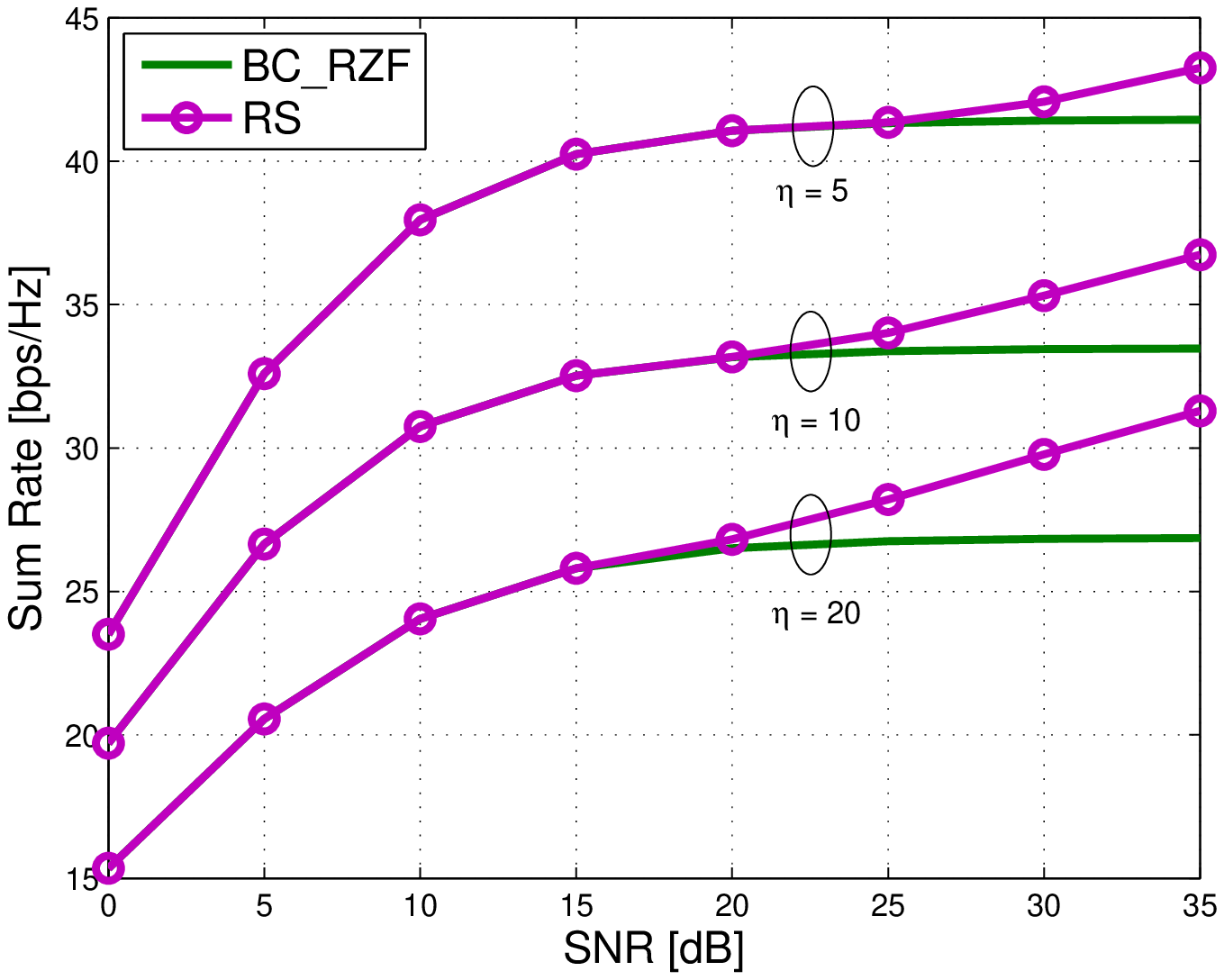}  } \hspace{20pt}
\subfigure[]{ \label{fig_RS4}  \includegraphics[width = 0.4\textwidth]{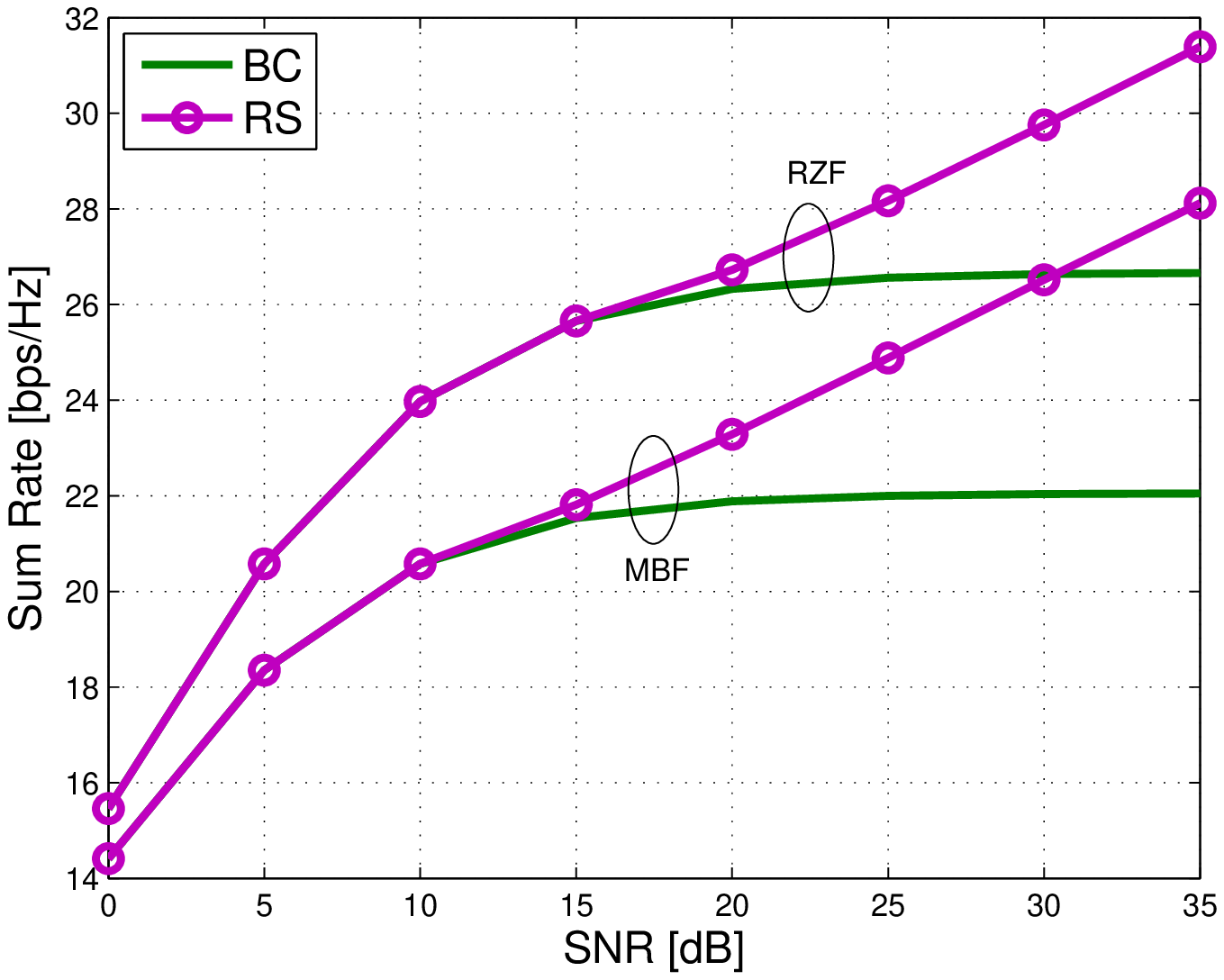} }
\caption{$\tau^2 = 0.5$. (a) Sum rate gain vs. $\eta = M/K$. (b) $K = 5$, RS with MBF vs. BC with RZF.}
\end{figure*}

In Fig. \ref{fig_RS3}, the sum rate gain of RS over BC\_RZF degrades with larger $K$ (smaller $\eta$), which confirms the discussion in Remark 2. Namely, the sum rate gain as well as the achievable rate of the common message becomes smaller as the number of users increases.

In general, a low-complexity precoder is desirable for massive MIMO systems \cite{herm2013}. MBF enjoys the lowest precoding complexity while ZF/RZF that achieves a much better performance than MBF involves complicated matrix inversion. Interestingly, RS enables to meet a certain sum rate requirement with a highly-reduced computational complexity compared with BC\_RZF. To identify the computational benefits of RS, Fig. \ref{fig_RS4} compares RS (MBF-precoded private messages) with RZF-precoded BC. The power splitting ratio $t$ of RS is computed via an exhaustive search\footnote{By calculating the asymptotic SINR of RS with MBF-precoded private messages, the closed-form $t$ can be obtained via \eqref{eq:poweralcrs2}.}. Recall that we assume a predefined set of $K$ user is scheduled, we observe that RS with MBF reaches the same rate performance as BC with RZF at SNR = 30 dB. In fact, RS simplifies the precoding design and decreases the computational complexity at the cost of an increased encoding and decoding complexity.

\subsection{HRS}
For simplicity, we assume $\tau_g = \tau, K_g = \bar{K}, b_g = \bar{b}, \forall g$. Consider $K = 12$ users equally clustered into $G = 4$ groups. We compare the proposed HRS scheme with the following baselines: \textbf{Baseline 1 (BC with two-tier precoder \cite{adhikary2013})}, \textbf{Baseline 2 (Baseline 1 with user scheduling at the group level)}: Within each group, a single user with the largest effective channel gain is selected and the precoder of the private message intended to each user is MBF. \textbf{Baseline 3 (Baseline 1 with user scheduling at the system level)}: User scheduling is performed at the system level such that the best user among all is selected. Two types of HRS are investigated: exhaustive search (HRS\_EXS) and closed-form (HRS\_CLF). Specifically, HRS\_EXS performs a simulation-based exhaustive search with step 0.01 for the best power splitting ratios $\alpha$ and $\beta$. HRS\_CLF allocates power by following the closed-form solution in Proposition 3.

\begin{figure*}[t] \centering
\subfigure[]{ \label{fig_AS1} \includegraphics[width = 0.4\textwidth]{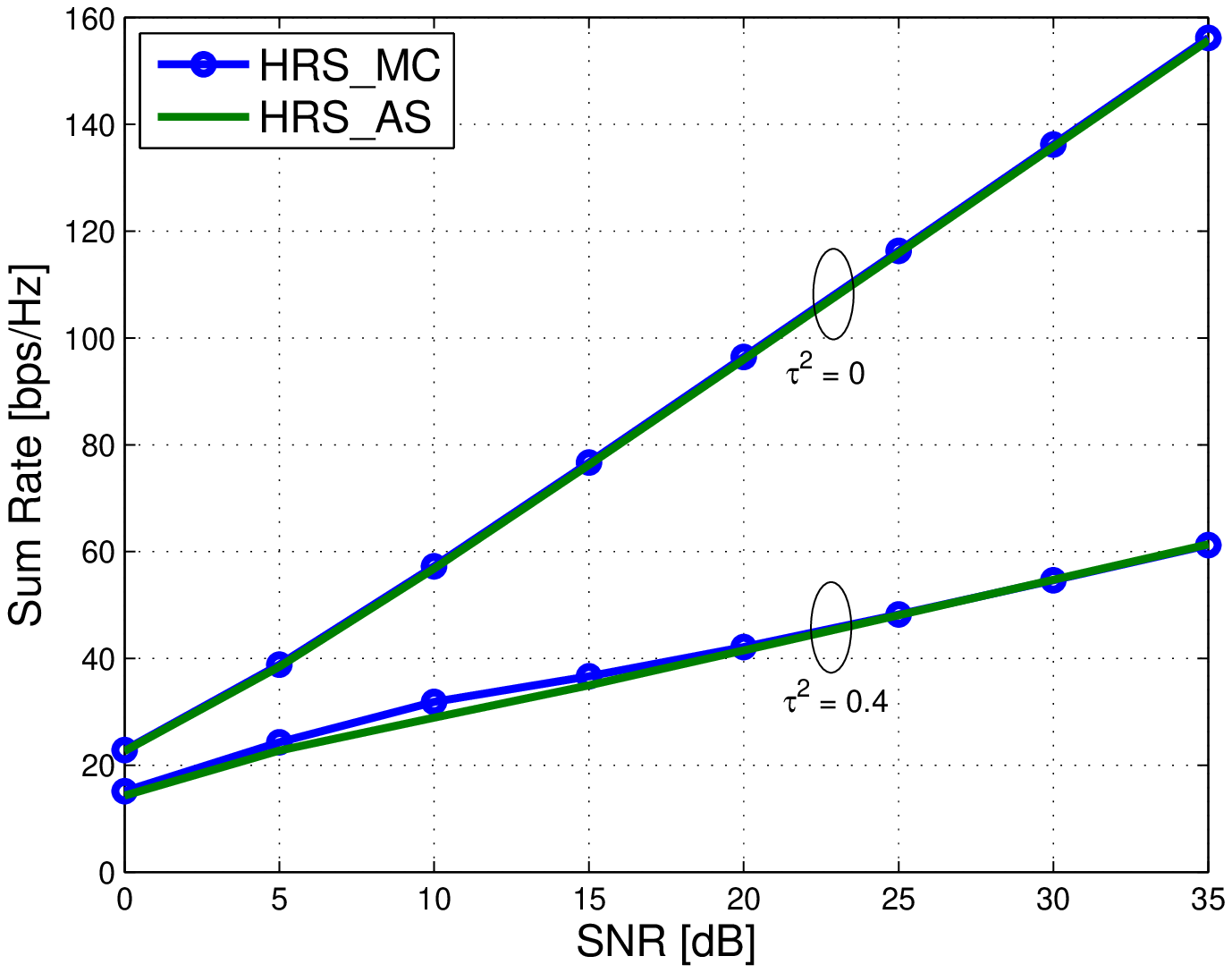}  } \hspace{20pt}
\subfigure[]{ \label{fig_AS2}  \includegraphics[width = 0.4\textwidth]{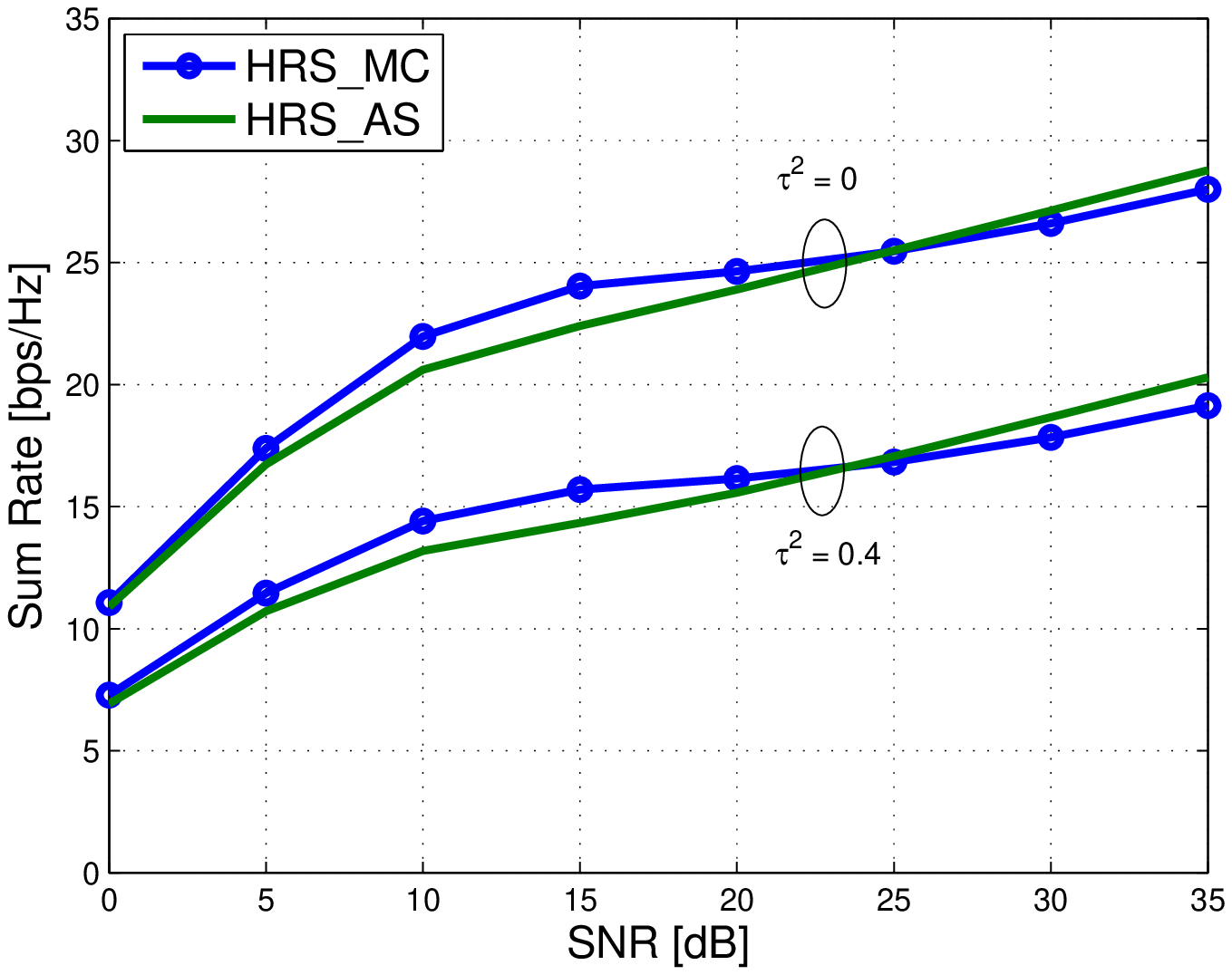} }
\caption{Asymptotic sum rate under perfect/imperfect CSIT. (a) disjoint eigen-subspace. (b) overlapping eigen-subspace.} \label{fig_AS}
\end{figure*}

\subsubsection{Validation of the Asymptotic Rate Analysis}

We compare the asymptotic sum rate \eqref{eq:sumratede} with simulations. Various CSIT qualities have been simulated and $\tau^2 = \{0, 0.4\}$ are taken as examples. For the outer precoder design, we set $\bar{b} = 15$ such that $\bar{K} \le \bar{b} \le M - (G-1) r^d$ and $\bar{b} \le r^d$, where $r^d = 20$\footnote{In this configuration, $r^d$ can be chosen as large as 25 due to $M = 100, G = 4$. Here, we set $r^d = 20$ since a relatively smaller $r^d$ enables to select stronger eigenmodes $\mathbf{F}^{(1)}_g$ thanks to the larger dimension of orthogonal subspace $M - \sum_{l \ne g} r^d$. In fact, simulation results revealed that HRS with $r^d = 20$ can achieve higher rate than that achievable with $r^d = 25$. We omit the simulation results for the sake of conciseness.} includes the dominant eigenvalues of $\mathbf{R}_g, \forall \, g$. To verify the effectiveness of the asymptotic sum rate approximation of HRS, we consider two scenarios with disjoint and overlapping eigen-subspaces, respectively. As an example, we set $\theta_g = -\frac{\pi}{2} + \frac{\pi}{3}(g-1)$ and $\Delta_g = \Delta = \frac{\pi}{8}, \forall \,g$ corresponding to disjoint eigen-subspaces $([\theta_g - \Delta_g, \theta_g + \Delta_g] \bigcap [\theta_l - \Delta_l, \theta_l + \Delta_l] = \emptyset, \forall \, l \ne g)$ while $\Delta_g = \Delta = \frac{\pi}{3}, \forall \,g$ leading to eigen-subspaces overlap.

In the scenario with disjoint eigen-subspaces, the inter-group interference is negligible and thereby the outer common message is unnecessary. When we further have perfect CSIT ($\tau^2 = 0$), there is no intra-group interference and the inner common messages are not needed. In this case, HRS boils down to two-tier precoding BC and Fig. \ref{fig_AS1} shows that the asymptotic sum rate {\small$R^{HRS, \circ}_{\scriptstyle{\text{sum}}}$} matches exactly the simulation result. As the CSIT quality decreases (increased $\tau^2$), inner common messages are exploited to mitigate intra-group interference. The asymptotic approximation \eqref{eq:sumratede} becomes less accurate but still valid to capture the asymptotic sum rate of HRS. By contrast, a larger angular spread generally leads to a larger rank of the spatial correlation matrix. With the outer precoder design as \eqref{eq:outerprecoder}, a large fraction of the power included in the weak eigenmodes is leaked into other groups, leading to strong inter-group interference. Fig. \ref{fig_AS2} indicates that the sum rate of HRS can be approximately characterized by \eqref{eq:sumratede}. In the low to medium SNR regime, HRS behaves as two-tier precoding BC where the rate gap between the asymptotic approximation and the simulation {\small$\frac{1}{K}(R^{HRS,\circ}_p - R^{HRS}_p) = \frac{1}{K}(R^{TTP, \circ}_{\scriptstyle{\text{sum}}} - R^{TTP}_{\scriptstyle{\text{sum}}})$} is within 0.2 bps/Hz. A similar behavior can be observed as well in \cite{wagner2012, Jaehyun2014}. At high SNR, the simulation {\small$R^{HRS}_p$} is around 1 bps/Hz lower than the asymptotic {\small$R^{HRS,\circ}_p$}, because the SINR of the outer common message {\small$\gamma^{oc} = \underset{g,k}{\min} \; \{\gamma^{oc}_{gk} \}$} is upper bounded by its asymptotic {\small$\gamma^{oc,\circ}$} for large but finite $M$. For example, when the precoder of the outer common message is designed such that all users experience the same {\small$\gamma^{oc,\circ}$}, the asymptotic approximation is then given by {\small$\gamma^{oc,\circ}$} while the simulated {\small$\gamma^{oc}$} is the minimum rate among all. It can be verified that this effect is mitigated when the channel hardens as the number of transmit antennas increases.

\begin{figure*}[t] \centering
\subfigure[]{ \label{fig_gain1} \includegraphics[width = 0.4\textwidth]{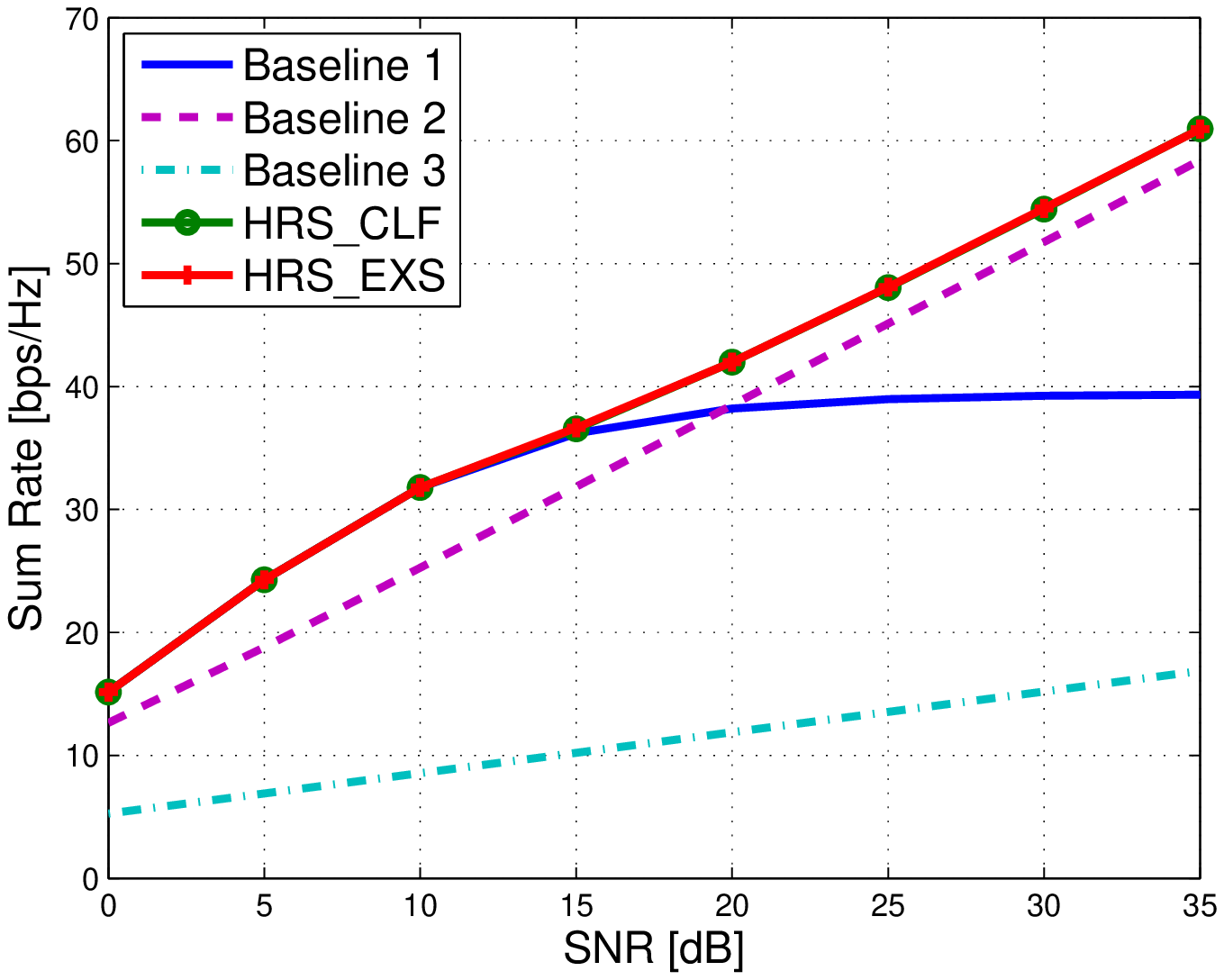}  } \hspace{20pt}
\subfigure[]{ \label{fig_gain2}  \includegraphics[width = 0.4\textwidth]{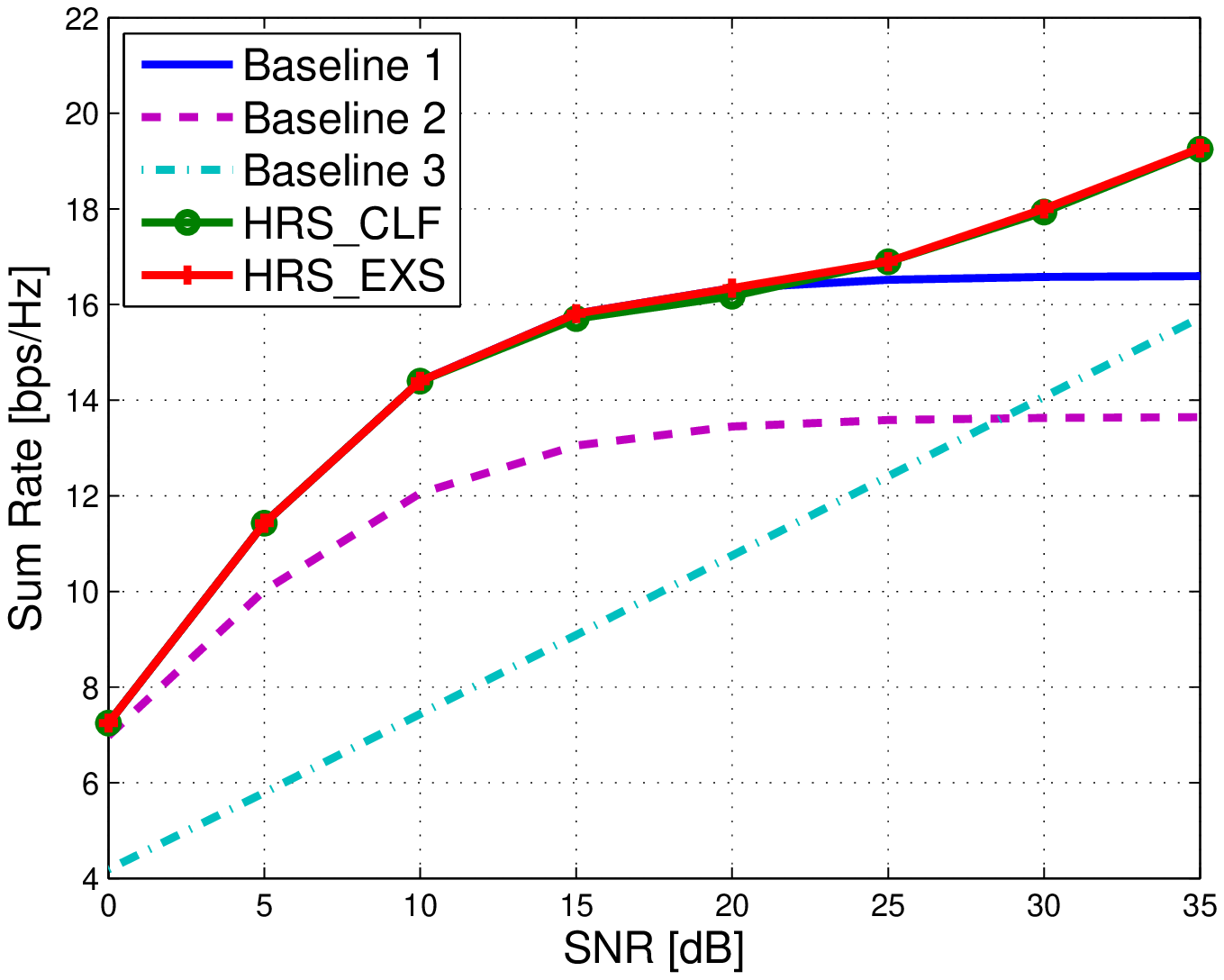} }
\caption{HRS vs. various baselines under imperfect CSIT. (a) disjoint eigen-subspace. (b) overlapping eigen-subspace.} \label{fig_gain12}
\end{figure*}

\subsubsection{Rate Performance Comparison}
Fig. \ref{fig_gain12} evaluates the benefits of HRS under imperfect CSIT ($\tau^2 = 0.4$) with the same system configuration as Fig. \ref{fig_AS}. With disjoint eigen-subspaces (negligible inter-group interference), Fig. \ref{fig_gain1} shows that conventional multiuser broadcasting scheme with two-tier precoder (Baseline 1) saturates at high SNR due to intra-group interference while user scheduling enables a multiplexing gain of 4 (Baseline 2) and 1 (Baseline 3), respectively. According to Proposition 3, HRS becomes a set of parallel inner RS. We observe that the proposed HRS scheme exhibits substantial rate gain over various baselines. For instance, the sum rate gain of HRS {\small$\Delta R^{HRS}$} over two-tier precoding BC at SNR = 30 dB is 15.5 bps/Hz. Based on \eqref{eq:hrsgain1} in Corollary 3.1, the asymptotic sum rate gain {\small$\Delta R^{HRS,\circ}$} is 19.5 bps/Hz. The 4 bps/Hz rate gap between {\small$\Delta R^{HRS}$} and {\small$\Delta R^{HRS,\circ}$} is explained as follows. The simulated rate loss {\small$R^{TTP}_{\scriptstyle{\text{sum}}} - R^{HRS}_p$} is indeed upper bounded by $G \log_2 (e)$ as \eqref{eq:hrsgaincal}. However, the simulated sum rate of the common messages {\small$R^{HRS}_{ic}$} is around 4 bps/Hz lower than {\small$R^{HRS,\circ}_{ic}$}. With severely overlapping eigen-subspaces (strong inter-group interference), HRS boils down to RS (with reduced-dimensional CSIT) at the system level according to Proposition 3, i.e., inner common messages are not transmitted. Fig. \ref{fig_gain2} reveals that the proposed HRS scheme outperforms two-tier precoding BC with/without user scheduling. The sum rate enhancement of HRS over two-tier precoding BC at SNR = 30 dB is 1.5 bps/Hz.

Interestingly, in both settings of Fig. \ref{fig_gain12}, the closed-form power allocation achieves almost the same sum rate as that of a simulation-based exhaustive search. This verifies the effectiveness of the power allocation strategy in Proposition 3. In Fig. \ref{fig_gain1} and Fig. \ref{fig_gain2}, respectively, we observe that the sum rate gain {\small$\Delta R^{HRS}$} of HRS over two-tier precoding BC increases by nearly $G$ and 1 bps/Hz for any 3 dB increment of power at high SNR, which verifies the discussion of Remark 3. In a nutshell, HRS exhibits robustness w.r.t. CSIT error and eigen-subspaces overlap. HRS behaves as two-tier precoding BC at low SNR, where the effect of imperfect CSIT/overlapping eigen-subspaces on the sum rate of broadcasting private messages is insignificant. At high SNR, by transmitting common message(s), the asymptotic multiplexing gain of HRS amounts to that of two-tier precoding BC with perfect user scheduling. Meanwhile, HRS exploits the rate benefits of two-tier precoding BC by transmitting the private messages with a fraction of the total power.


Comparing baseline 1 and 2, we see the importance of user scheduling for multiuser broadcasting strategy. User scheduling would obviously be useful to HRS as well, but HRS shows a very competitive performance even without user scheduling. This is particularly attractive in massive MIMO where the number of users $(K)$ can potentially be large. Hence HRS would decrease the burden on the scheduler and the precoder design but increases the complexity of the encoding and decoding strategies.

\subsection{RS vs. HRS}

\begin{figure}[t]
  \centering
  \includegraphics[width = 0.4\textwidth]{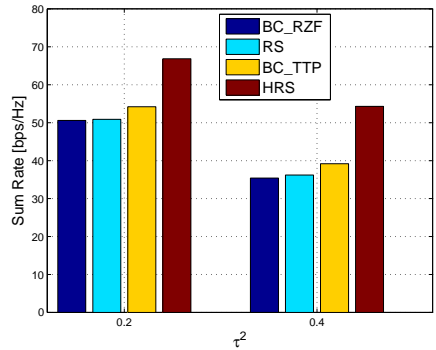}\\
  \caption{Sum rate comparison with various CSIT qualities, SNR = 30 dB.}\label{fig_gain5}
\end{figure}

To examine the suitability of HRS in spatially correlated massive MIMO, we compare HRS with BC\_RZF/RS with full-dimensional CSIT in Section \ref{RS} and two-tier precoding BC with reduced-dimensional CSIT for various CSIT qualities. We assume the same system configuration as Fig. \ref{fig_gain1}. Overall, Fig. \ref{fig_gain5} shows that a lower CSIT quality (i.e., larger $\tau^2$) degrades the rate performance of these schemes.

In Fig. \ref{fig_gain5}, we take $\tau^2 = 0.4$\footnote{When the imperfect CSIT is due to the quantization errors, the number of feedback bits required to achieve a certain CSIT quality is proportional to the dimension of the quantized channel. Then, the CSIT quality of two-tier precoding BC/HRS would be better than that of one-tier precoding BC/RS for a given number of feedback bits. In this case, the rate gap between one-tier and two-tier precoding strategies will be even larger. Nevertheless, we assume the same CSIT quality for simplicity of exposition.} as an example. The sum rate of RS slightly outperforms BC\_RZF around 1 bps/Hz. Recall that the achievable rate of the common message in RS is the minimum rate among all users $(K = 12)$, the sum rate gain from transmitting a common message appears small at SNR = 30 dB. By contrast, the sum rate gain of HRS over BC\_TTP is 15.5 bps/Hz. The large gain of HRS is enabled by multiple inner common messages while the achievable rate of each inner common message is the minimum rate among a smaller number of users $(\bar{K} = 3)$. This observation confirms that if the channel second-order statistics are available to the transmitter, the proposed HRS scheme is better suited for massive MIMO deployments than than a simple RS strategy.

Moreover, with reduced-dimensional CSIT, BC with two-tier precoder (BC\_TTP) and HRS achieve much higher rates than BC\_RZF and RS. It is because the outer precoder exploiting the long-term CSIT partitions users into a set of non-interfering groups and each user experiences interference from fewer users compared with BC\_RZF and RS.

\section{CONCLUSION} \label{conclusion}

Due to imperfect CSIT, the rate performance of conventional multiuser broadcasting schemes is severely degraded. To tackle the multiuser interference, a Rate-Splitting approach has been proposed. RS packs part of one selected user's message into a common message that can be decoded by all users and superimposes the common message on top of the private messages. We generalized the RS method to the large-scale array regime. By further exploiting the channel second-order statistics and a two-tier precoding structure, we proposed a novel Hierarchical-Rate-Splitting strategy. Particularly, on top of the private messages, HRS transmits an outer common message and multiple inner common messages that can be decoded by all users and a subset of users, respectively. The outer common message tackles the inter-group interference due to overlapping eigen-subspaces while the inner common messages helps with mitigating the intra-group interference due to imperfect CSIT.

For RS and HRS, we derived the precoder design, asymptotic rate performance and power allocation. Interestingly, to meet a certain sum rate requirement, RS highly decreases the complexity of precoder design and scheduling at the expense of an increase in complexity of the encoding and decoding strategy. Moreover, simulation results showed that the rate performance of the conventional broadcasting schemes saturates at high SNR due to imperfect CSIT and the sum rate gain of RS over BC with RZF increases with the available transmit power. When users in different groups have disjoint eigen-subspaces, the sum rate gain of HRS over BC with two-tier precoder is much larger than the gain achievable with RS. In a nutshell, RS and HRS exhibit robustness w.r.t. CSIT error and/or eigen-subspaces overlaps while HRS is a general and particularly suited framework for the massive MIMO deployments.

We believe that the idea of RS and HRS has a huge potential in a wide range of wireless communication systems including massive MIMO and single-cell/multi-cell multiuser MIMO broadcast channel, operated at microwave/millimeter wave with FDD/TDD mode, equipped with sufficient/limited radio frequency (RF) chains at the BS. In practice, CSIT is imperfect due to imperfect channel estimation, limited feedback, antenna miscalibration, pilot contamination, etc. The effectiveness of RS and HRS have been demonstrated to mitigate the resultant interference.

\appendix

\subsection{Proof of Proposition 2} \label{sec:prop2}
When {\small$\frac{Pt}{K} \left(\xi^{\circ}\right)^2 \Upsilon^{\circ}_{k} \Omega_{k} > 1$}, the equality in \eqref{eq:rs_pow1} is nearly established, i.e., the private messages of RS achieve approximately the same sum rate as the conventional multiuser BC with full power. The power splitting ratio $t$ is then designed as {\small$\frac{Pt}{K} \left(\xi^{\circ}\right)^2 \Upsilon^{\circ}_{k} \Omega_{k} = K$}, i.e., {\small$t = {K}/{(P \, \Upsilon^{\circ}_{k} \Omega_{k}/\Psi^{\circ})}$}. The rationale behind this design is two-fold. On the one hand, the number of users $K$ is generally much larger than 1, which leads to an asymptotically tight approximation. On the other hand, the achievable rate of the common message decreases as $K$ increases due to minimum constraint. This effect can be observed via $\eta = M/K$ in the asymptotic $\gamma^{c,\circ}$. Then, the power allocated to the common message $P(1 - t)$ should be reduced as $K$ becomes larger. Otherwise, suppose $P(1 - t)$ is constant independent of $K$. As $K$ increases in \eqref{eq:rsgain}, the achievable rate of the common message {\small$\log_2 (1 + \gamma^{c,\circ})$} cannot compensate the loss {\small$\sum^K_{k = 1} \big(\log_2 (1 + \gamma^{p,\circ}_{k}\big) - \log_2 \big(1 + \gamma^{RZF,\circ}_{k}) \big)$} incurred from the above approximation. Moreover, $\Omega_k$ in \eqref{eq:rs_desub3} can be approximated by $\tau^2$ and the approximation is tight when $m^{\circ}_k$ is large. Thus, {\small$t = K/(P \Gamma_k)$}, where {\small$\Gamma_k = ( \Upsilon^{\circ}_{k}\, \tau_k^2) /\Psi^{\circ}$}. To establish the equality \eqref{eq:rs_pow1} for $\forall k$, the power splitting ratio $t$ is chosen as the largest one. We then obtain \eqref{eq:poweralcrs2} by truncating $t$ at 1 wherever applicable.

At low SNR, $t = 1$ from \eqref{eq:poweralcrs2} turns RS into BC and leads to {\small$\Delta R^{RS,\circ} = 0$}. Namely, broadcasting multiple private messages is operated in the non-interference limited SNR regime and thereby a common message is unnecessary. At high SNR, $t < 1$ indicates that we transmit a common message with remaining power beyond the saturation of the private message transmission. Due to power reduction to the private messages, we first upper bound the rate loss {\small$R^{RZF,\circ}_{\scriptstyle{\text{sum}}} - R^{RS,\circ}_p$}
\begin{equation} \label{eq:rsgaincal}
\begin{array}{lcl}
&=& \sum^K_{k = 1} \Big(\log_2 \big(1 + \frac{S}{P \Gamma_k+ 1} \big) - \log_2 \big(1 + \frac{S}{P \Gamma_k + \frac{1}{t}} \big) \Big)\\
&=& \sum^K_{k = 1} \big(\log_2 (S + P \Gamma_k + 1) - \log_2 (S + P \Gamma_k + \frac{1}{t}) \\ && \qquad \qquad + \log_2 (P \Gamma_k + \frac{1}{t}) - \log_2 (P \Gamma_k + 1 )\big)\\
&\overset{(a)}{\le}& \sum^K_{k = 1} \Big(\log_2 (P \Gamma_k  + \frac{1}{t}) - \log_2 (P \Gamma_k + 1 \big)\Big) \\
&\overset{(b)}{\le}& \sum^K_{k = 1} \big(\log_2 (1 + \frac{1}{K}) - \log_2 (1 + \frac{1}{P \Gamma_k})\big) \\
&\overset{(c)}{\le}& K\log_2(1 + 1/K)
\overset{(d)}{\le} \log_2 (e),
\end{array}
\end{equation}
\vspace{-2pt} \hspace{-10pt}
where {\small$S = \frac{P}{K} \left(\xi^{\circ}\right)^2 \Phi_k$. $(a)$} is obtained since $1/t \ge 1, \forall t \in (0, \,1]$. By replacing $t = K/{(P \Gamma)}$ with {\small$\Gamma = \underset{k}{\text{min}} \, \{\Gamma_k\}$} by $t = K/{(P \Gamma_k)}$, $(a)$ is lower bounded as $(b)$. Removing {\small$\log_2 (1 + \frac{1}{P \Gamma_k})$}, we have $(c)$ which is tight at high SNR. $(d)$ is obtained since $K\log_2(1 + 1/K) \in [1, \log_2 (e))$ is an increasing function of $K$ for $K \ge 1$. By plugging $(d)$ into \eqref{eq:rsgain}, we then obtain \eqref{eq:rsgain2}.

\subsection{Proof of Proposition 3} \label{sec:prop3}

The inter-group interference is captured by {\small$\bar{\mathbf{R}}_{gl} = \mathbf{B}^H_l \mathbf{R}_g \mathbf{B}_l,\; \forall g \ne l$}. Let us firstly consider the weak inter-group interference case, i.e., {\small$\bar{\mathbf{R}}_{gl} \approx \mathbf{0}_{b' \times b'}$} and further {\small$\Upsilon^{\circ}_{gl} \approx 0, \; \forall g \ne l$}, i.e., the inter-group interference is sufficiently small and therefore can be negligible. The sum rate of the private messages transmission is limited by intra-group interference. Based on \eqref{eq:desnr2} $\sim$ \eqref{eq:desnr4} and \eqref{eq:sumratede}, the outer common message suffers from more interference while contributing less rate than the inner common messages, since the achievable rate of the outer common message has a pre-log factor of 1 which is smaller than that of the inner common messages ($G > 1$). The optimal $\beta$ that maximizes the sum rate of HRS \eqref{eq:sumratede} is $\beta = 1$. Then, \eqref{eq:desnr4} and \eqref{eq:snrjsdm} become
\begin{equation} \label{eq:prop1}
\gamma^{p,\circ}_{g} = \frac{\alpha_g \frac{P}{K} \big(\xi^{\circ}_{g} \big)^2 \Phi_g}{ \alpha_g \big(\xi^{\circ}_{g} \big)^2 \Upsilon^{\circ}_{gg} \Omega_g + 1}, \quad
\gamma^{BC, \circ}_{g} = \frac{\frac{P}{K} \big(\xi^{\circ}_{g} \big)^2 \Phi_g}{ \big(\xi^{\circ}_{g} \big)^2 \Upsilon^{\circ}_{gg} \Omega_g + 1}.
\end{equation}

Substituting \eqref{eq:prop1} into \eqref{eq:poweralc1}, the equality is approximately established when {\small$\alpha_g (\xi^{\circ}_{g} )^2 \Upsilon^{\circ}_{gg} \Omega_g > 1$}. Following a similar philosophy of $t$ in Proof of Proposition 2, the intra-group power splitting ratio is designed as {\small$\alpha_g = {K_g}/{\big((\xi^{\circ}_{g} )^2 \Upsilon^{\circ}_{gg} \Omega_g }\big)$}. Otherwise, as $K_g$ increases, the achievable rate of the inner common message {\small$\log_2 (1 + \gamma^{c,\circ}_g)$} cannot compensate the loss {\small$K_g \big(\log_2 (1 + \gamma^{p,\circ}_{g}) - \log_2 (1 + \gamma^{TTP,\circ}_{g} ) \big)$} incurred from the above approximation. Based on \eqref{eq:desub1} $\sim$ \eqref{eq:desub5}, $\alpha_g$ is determined by
\begin{equation} \label{eq:prop2}
\big(\xi^{\circ}_{g} \big)^2 \Upsilon^{\circ}_{gg} \Omega_g = \frac{P K_g}{K} \frac{\text{tr}\big(\bar{\mathbf{R}}_{gg}\mathbf{T}_{g} \bar{\mathbf{R}}_{gg} \mathbf{T}_{g}\big)}{\text{tr}\big(\bar{\mathbf{R}}_{gg}\mathbf{T}^2_{g}\big)} \Omega_g.
\end{equation}

In order to obtain a more insightful understanding of the effects of system parameters, we consider a high SNR approximation of \eqref{eq:prop2}. At high SNR ($\varepsilon \approx 0$), the RZF matrix in \eqref{eq:innerprecoder} converges to the ZF matrix. From \cite[Theorem 3]{wagner2012}, $\mathbf{T}_g$ in \eqref{eq:desub5} becomes
\begin{equation} \label{eq:prop3}
\mathbf{T}_g = \left( \frac{K_g}{b_g} \frac{\bar{\mathbf{R}}_{gg}}{m^{\circ}_{g}} + \mathbf{I}_{b_g} \right)^{-1} \approx \left( \frac{K_g}{b_g} \frac{\bar{\mathbf{R}}_{gg}}{m^{\circ}_{g}} \right)^{-1},
\end{equation}
where $\bar{\mathbf{R}}_{gg}$ is a diagonal matrix from \eqref{eq:outerprecoder}. Since $\mathbf{B}_g$ lies in the dominant eigenmodes of $\mathbf{R}_g$, the diagonal elements of $\bar{\mathbf{R}}_{gg}$ are much larger than 1 and therefore the approximation in \eqref{eq:prop3} is feasible. Moreover, we have {\small$\Omega_g \approx \frac{K_g-1}{K_g} \tau_g^2$} due to the fact $m^{\circ}_{g} \ge 1$ in the asymptotic $M$ regime. Plugging \eqref{eq:prop3} into \eqref{eq:prop2} leads to {\small$\big(\xi^{\circ}_{g} \big)^2 \Upsilon^{\circ}_{gg} \Omega_g \approx  \frac{P}{K} \frac{b_g (K_g-1)}{\text{tr}\big(\bar{\mathbf{R}}^{-1}_{gg}\big)} \tau_g^2$}. Since the same $\alpha$ is applied to all groups, we choose the largest one to guarantee \eqref{eq:poweralc1}:
\begin{equation} \label{eq:prop5}
\alpha = \frac{K_g}{P \cdot \Gamma_{IG}}, \; \Gamma_{IG} = \underset{g}{\text{min}} \; \bigg\{ \frac{\tau_g^2}{K} \frac{b_g (K_g - 1)}{\text{tr}\big(\bar{\mathbf{R}}^{-1}_{gg}\big)} \bigg\}.
\end{equation}

Secondly, consider the case with {\small$\sum_{l \neq g} (\xi^{\circ}_{l})^2 \Upsilon^{\circ}_{gl} > (\xi^{\circ}_{g} )^2 \Upsilon^{\circ}_{gg}$}. Since {\small$\Omega_g < 1$} from \eqref{eq:desub4}, the sum rate of the private messages based on \eqref{eq:snrjsdm} is dominated by inter-group interference. Substituting \eqref{eq:desnr4}, \eqref{eq:snrjsdm} into \eqref{eq:poweralc1}, the equality is approximately established when {\small$\beta \sum_{l \neq g} \left(\xi^{\circ}_{l}\right)^2 \Upsilon^{\circ}_{gl} + \beta \alpha \big(\xi^{\circ}_{g} \big)^2 \Upsilon^{\circ}_{gg} \Omega_g > 1$} and $\alpha = 1$. Following a similar philosophy of $t$ in Proof of Proposition 2, the inter-group power splitting ratio can be designed as {\small$\beta_g = {K}/{\big(\sum_{l \neq g} (\xi^{\circ}_{l})^2 \Upsilon^{\circ}_{gl} + (\xi^{\circ}_{g} )^2 \Upsilon^{\circ}_{gg} \Omega_g }\big)$}. However, we adopt a conservative design of $\beta_g$ as
\begin{equation} \label{eq:prop6}
\beta_g = \frac{K}{\sum_{l \neq g} \left(\xi^{\circ}_{l}\right)^2 \Upsilon^{\circ}_{gl} + K_g } \ge \frac{K}{\sum_{l \neq g} \left(\xi^{\circ}_{l}\right)^2 \Upsilon^{\circ}_{gl} + \big(\xi^{\circ}_{g} \big)^2 \Upsilon^{\circ}_{gg} \Omega_g },
\end{equation}
which is due to the fact that {\small$(\xi^{\circ}_{g})^2 \Upsilon^{\circ}_{gg} \Omega_g > K_g$} at high SNR (interference regime). The rationale behind this conservative design is two-fold. Larger $\beta$ is more capable to maintain \eqref{eq:poweralc1}. Besides, it enables a distributed design of power allocation, i.e., $\beta$ is determined only by the long-term inter-group interference. By plugging \eqref{eq:desub1} $\sim$ \eqref{eq:desub32} and \eqref{eq:prop3} into \eqref{eq:prop6} and denoting $\beta$ as the largest $\beta_g$, we have
\begin{equation} \label{eq:prop7}
\beta = \frac{K}{P \cdot \Gamma_{OG} + K_g}, \; \Gamma_{OG} =  \underset{g}{\text{min}} \; \bigg\{ \sum_{l \ne g} \frac{K_g}{K} \frac{\text{tr}\big(\bar{\mathbf{R}}_{gl} \bar{\mathbf{R}}^{-1}_{ll}\big)}{\text{tr}\big(\bar{\mathbf{R}}^{-1}_{ll}\big)} \bigg\}
\end{equation}

Since $0 < \alpha, \beta \le 1$, we assume implicitly that $\forall \alpha, \beta > 1$ is truncated at 1 wherever applicable.

\subsection{Proof of Corollary 3.1} \label{sec:coro3}
We here provide a sketch proof, since it follows a similar philosophy of {\small$\Delta R^{RS,\circ}$} in Proof of Proposition 2. In the weak inter-group interference regime, $\beta = 1$ from \eqref{eq:poweralc2}. We first upper bound the rate loss {\small$R^{TTP,\circ}_{\scriptstyle{\text{sum}}} - R^{HRS,\circ}_p$} at high SNR
\begin{equation} \label{eq:hrsgaincal}
\begin{array}{lcl}
&=& \sum^G_{g = 1} K_g \Big(\log_2 \big(1 + \frac{S_g}{\Gamma_g+ 1} \big) - \log_2 \big(1 + \frac{S_g}{\Gamma_g + \frac{1}{\alpha}} \big) \Big)\\
&\le& \sum^G_{g = 1} K_g \log_2 (1 + 1/K_g) \le G \log_2 (e),
\end{array}
\end{equation}
where {\small$S_g = \frac{P}{K} \big(\xi^{\circ}_{g} \big)^2 \Phi_g$} and {\small$\Gamma_g = (\xi^{\circ}_{g} )^2 \Upsilon^{\circ}_{gg} \Omega_g$}. The sum rate gain {\small$\Delta R^{HRS,\circ}$} is lower bounded as \eqref{eq:hrsgain1}. In the strong inter-group interference regime, the rate loss is upper bounded as
\vspace{-15pt}
\begin{equation} \label{eq:hrsgaincal2}
\begin{array}{lcl}
&=& \sum^G_{g = 1} K_g \Big(\log_2 \big(1 + \frac{S_g}{\Gamma_g+ 1} \big) - \log_2 \big(1 + \frac{S_g}{\Gamma_g + \frac{1}{\beta}} \big) \Big)\\
&\le& \sum^G_{g = 1} K_g \log_2 (1 + 1/K) \\
&=& K \log_2 (1 + 1/K)  \le \log_2 (e),
\end{array}
\end{equation}
where {\small$\Gamma_g = \sum_{l \neq g} \left(\xi^{\circ}_{l}\right)^2 \Upsilon^{\circ}_{gl} + \big(\xi^{\circ}_{g} \big)^2 \Upsilon^{\circ}_{gg} \Omega_g$}. Then, the sum rate gain is lower bounded as \eqref{eq:hrsgain2}.

\begin{spacing}{0.98}
\bibliographystyle{IEEEtran}
\bibliography{reference}
\end{spacing}

\end{document}